\documentstyle[aps,twocolumn,prb,epsfig]{revtex}

\begin{document}
\draft
\title{Ferromagnetic transition and phase diagram of the one-dimensional 
Hubbard model with next-nearest-neighbor hopping}

\twocolumn[ 
\hsize\textwidth\columnwidth\hsize\csname@twocolumnfalse\endcsname 

\author{S.\ Daul and R. M.\ Noack}
\address{
 Institut de Physique Th\'{e}orique,
Universit\'{e} de Fribourg, CH-1700 Fribourg, Switzerland}
\maketitle

\begin{abstract}

We study the phase diagram of the one-dimensional Hubbard model
with next-nearest-neighbor hopping using exact
diagonalization, the density-matrix renormalization group, the Edwards
variational ansatz, and an adaptation of weak-coupling calculations
on the two-chain Hubbard model.
We find that a substantial region of the strong-coupling phase diagram
is ferromagnetic, and that three physically different limiting cases
are connected in one ferromagnetic phase.
At a point in the phase diagram at which there are two
Fermi points at weak coupling, we study carefully the phase transition
from the paramagnetic state to the fully polarized one as a function of
the on-site Coulomb repulsion.
We present evidence that the transition is second order and determine
the critical exponents numerically.
In this parameter regime, the system can be
described as a Luttinger liquid at weak coupling. 
We extract the Luttinger-liquid parameters and
show how their behavior differs from that of the nearest-neighbor
Hubbard model.
The general weak-coupling phase diagram can be mapped onto that of the
two-chain Hubbard model.
We exhibit explicitly the adapted phase diagram and determine its
validity by numerically calculating spin and charge gaps using the
density-matrix renormalization group.

\end{abstract}
\pacs{PACS Numbers: 71.10.Fd, 75.10.Lp and 75.40.Mg}
]

\section{INTRODUCTION}

Unlike strongly correlated phenomena such as antiferromagnetism or
superconductivity which can be treated starting from a weak-coupling
point of view, metallic ferromagnetism is an intrinsically
intermediate or strong coupling phenomenon.
Because of this, the origin of metallic ferromagnetism is still poorly
understood, even after decades of research.
The simplest model of correlated electrons, 
the Hubbard model, was introduced simultaneously by Gutzwiller,
\cite{Gutzwiller63} Hubbard \cite{Hubbard63} and Kanamori
\cite{Kanamori63} in  1963 in order to study ferromagnetism. 
Indeed, at a mean-field level, the Hubbard model seems to be a good
starting point for ferromagnetism, because the Stoner criterion
predicts a ferromagnetic ground state for a wide range of parameters. 
However, the inclusion of correlation effects makes the
conditions for the appearance of ferromagnetism much more restrictive.
\cite{Hubbard63}

There is one limit of the Hubbard model in which a
ferromagnetic state can be obtained within an exact treatment.
For $U=\infty$ the ground state of the
half-filled system has macroscopic degeneracy since all states
with different spin $S$ have the same energy.
For bipartite lattices (such as the hypercubic and bcc lattices) in
dimension $d \ge 2$, and for fcc and hcp lattices with negative
hopping integrals, Nagaoka \cite{Nagaoka66} proved that when
one hole is then added to the $L$-site system, the ground state of the
model has maximum spin.
It has not yet been proven possible to generalize Nagaoka's
proof for the stability of ferromagnetism to a finite density of holes 
$\delta=1-n > 0$, making the treatment of the thermodynamic limit
problematic.
One can, however, use the opposite approach and try to show that
the ferromagnetic state is unstable by applying a suitable variational
wave function. 
Starting from the fully polarized state, one can flip a spin in an
appropriate manner and then see if the corresponding energy is lower
than the ferromagnetic one.
If a lower energy can be found, the fully polarized ferromagnetic
state is proven to be unstable.
Recently, by choosing a highly sophisticated variational state, 
Wurth and co-workers \cite{Wurth95} were able to
bring the critical hole density $\delta_c$ above which the Nagaoka
state is unstable at $U=\infty$ down to $\delta_c = 0.251$ for a
square lattice.
Exact diagonalization \cite{HirschThesis} and DMRG \cite{Liang95}
studies suggest that $\delta_c$ could be even lower. 
Hence it cannot be ruled out that $\delta_c=0$ for bipartite
lattices, as is the case for the hyper-cubic lattice in infinite 
dimensions.\cite{Fazekas90} 
In contrast, for non-bipartite lattices, a partly polarized ground
state (ferrimagnetism) has been obtained by Lieb. \cite{Lieb89}

For one-dimensional systems, the situation is even less favorable for 
ferromagnetism. 
Lieb and Mattis \cite{Liebmattis62} have proven that the ground
state is unmagnetized for any real and particle-symmetric but
otherwise arbitrary interaction. 
This theorem applies to a single band in $d=1$, provided that
the hopping is only between nearest neighbors and the interaction
involves only densities. 
Since both conditions are fulfilled in the  Hubbard model,
its ground state in $d=1$ cannot be ferromagnetic.
 
In principle, the Hubbard model is obtained from an extreme truncation
of a more general Hamiltonian describing interacting electrons in a solid. 
Only the on-site interaction and one relevant band are kept.
Nearest-neighbor interaction (e.g. direct exchange), band degeneracy
and the associated Hund's rule couplings are totally neglected.
In addition, the non-interacting band structure and density of states
can be strongly affected by the lattice structure.
In order to enhance the possibility of ferromagnetism, one can modify
the simple Hubbard model by putting some of these neglected features
back in.

Strack and Vollhardt \cite{Strack94} have studied a Hubbard model to 
which they have added all possible nearest-neighbor interactions:
the usual interaction between charge ($V$), the density-dependent
hopping ($X$), the Heisenberg exchange ($F$) and pair hopping ($F'$).
They show that for a particular range of parameter values the model
has maximum total spin at half-filling. 
These arguments can also be extended to the Nagaoka case of one hole. 
\cite{Kollar96}

Another option is to take a multi-band Hamiltonian with $d$-band
degeneracy together with a Hund's rule coupling between the different
$d$ orbitals. Okabe \cite{Okabe97} has investigated 
the stability of the ferromagnetic state of such a model variationally,
while Fleck and coworkers have studied a similar model including
next-nearest-neighbor hopping. \cite{Fleck96}
These authors claim that the Hund's rule coupling is necessary to
obtain ferromagnetism.
Very recently, B\"unemman and coworkers 
\cite{Bunnemann97} have studied a two-band Hubbard model with a multi-band 
Gutzwiller wave function. They found that a ferromagnetic transition occurs
at large interaction and stress that a finite value of the exchange
interaction is also required.
We will see here that ferromagnetism can be obtained in a
non-orbitally-degenerate model. 

Mielke \cite{Mielke93} has proven the following theorem, equivalent to Hund's 
rule, for a Hubbard model with a flat band. 
If the model has an $M$-fold degenerate single-particle
ground state, then for any number of electrons $N \leq M$ the fully
polarized state (with total spin $S=\frac{N}{2}$) is a
ground state of the system.
Additional conditions that determine whether this ground state is unique
are also given.
This theorem has also been extended to nearly flat bands. \cite{MielkeTasaki93}
Recently, Tasaki \cite{Tasaki95} has considered a two-band Hubbard model
with next-nearest-neighbor hopping. 
He has proven that the quarter--filled system 
(average electron density $n=0.5$) is ferromagnetic for large enough
on-site interaction $U$.
Penc and coworkers \cite{Penc96} have extended this result to other 
fillings by studying a Hamiltonian in which a chemical potential is
added to all even sites of the lattice to make a perturbative argument
valid.
We shall see below that this term is unnecessary to obtain a ferromagnetic 
ground state.

Several authors have studied the ordinary Hubbard model on various lattice
structures. 
Ulmke \cite{Ulmke97} investigated the case of an fcc lattice
using dynamical mean-field theory and quantum Monte Carlo simulations. 
He found ferromagnetism for intermediate values of $U$, using the
density of states of the infinite-dimensional system. 
For the three-dimensional density of states he
found that one must add next-nearest-neighbor hopping in order to obtain
ferromagnetism. 
His general conclusion is that a necessary condition for ferromagnetism
is a density of states with large spectral weight near 
the lower band edge.
Hanisch and coworkers \cite{Hanisch97} have investigated the stability of 
saturated ferromagnetism using a variational approach for various lattice 
structures in two and three dimensions.
Their conclusions are similar to the previous ones, namely that a
particle-hole asymmetry and a divergent density of states at the lower
band energy are necessary
ingredients for obtaining a ferromagnetic ground state.
Similar conclusions have been reached very recently using the
spectral density approach \cite{Hermann97} and dynamical mean-field theory
with the Non-Crossing Approximation. \cite{Obermeier97}

In the small density limit,
the ferromagnetic state of the Hubbard model with
arbitrary non-diagonal hopping and with a band structure
with a quadratic dispersion about the band minima
has been shown to be unstable for $d>3$,
while for $d=2$ a small window of parameters for which the fully
polarized state is not destabilized still remains. \cite{Pieri96}
Indeed, a projector quantum Monte Carlo calculation
\cite{Hlubina97} yields ferromagnetism precisely in this allowed
region for the two-dimensional Hubbard model on a square lattice with
next-nearest-neighbor hopping. 
Finally, a renormalization group calculation for this model
\cite{Alavarez97} also yields ferromagnetism in a particular regime.

In this work, \cite{Daulthesis} we study perhaps the simplest case of a Hubbard 
model exhibiting ferromagnetism: the one-dimensional
Hubbard model with an additional next-nearest-neighbor hopping.
Previously, exact diagonalization, \cite{Pieri96} variational
\cite{DaulPieri97} and Density-Matrix Renormalization Group (DMRG)
calculations \cite{DaulNoack97} on
this model have already concluded that there is an extensive ferromagnetic
phase for large enough coupling $U$.
A weak-coupling analysis \cite{Fabrizio96} applied to this
one-dimensional model leads to a phase with a spin gap which is the
one-dimensional analog of a superconductor.
Projector QMC and DMRG calculations for the special case of 
half-filling \cite{Kuroki97} have recently been carried out at
weak to intermediate coupling and are consistent with the weak-coupling
analysis. 
Here we treat the strong and weak-coupling phase diagrams of
the model comprehensively with numerical and variational techniques,
and link them by studying the phase transition using exact
diagonalization and DMRG calculations. 

This paper is organized as follows.
In Sec.\ \ref{modelprops}, we discuss the basic properties of the model,
motivate the existence of ferromagnetism by discussing calculations on
three and four site clusters, and discuss exactly treatable limiting
cases at interaction $U=\infty$, some of which yield ferromagnetic
ground states.
The results of a variational calculation of the $U=\infty$ phase
diagram using the Edwards variational ansatz are presented in 
Sec.\ \ref{variational}.
While this technique was previously applied to this model in 
Ref.\ \onlinecite{DaulPieri97} at finite $U$, here we discuss
the $U=\infty$ phase diagram and include a determination of the total
spin of the variational state.
Exact diagonalization calculations, presented in 
Sec.\ \ref{exactdiag}, are used to illustrate the determination of the
critical interaction strength $U_c$ at the ferromagnetic transition,
to determine the order of the phase transition, and to examine the
scaling of $U_c$ for small next-nearest-neighbor hopping $t_2$.
In Sec.\ \ref{DMRG} we use the DMRG to determine $U_c$ as a function
of density for three different $t_2$ regimes, discuss the behavior of
the Luttinger liquid parameters at the transition from a Luttinger
liquid regime to the ferromagnetic regime, and discuss the behavior of
the spin-spin correlation function near the transition.
We present a determination of the strong-coupling ($U=\infty$) and
weak-coupling phase diagrams in Sec.\ \ref{phasediagrams}.
The strong-coupling phase diagram is calculated using the DMRG, and
the weak-coupling phase diagram is determined by adapting the results
of Balents and Fisher\cite{Balents96} for the two-chain Hubbard model to
this model, as suggested by Fabrizio. \cite{Fabrizio96}
A DMRG calculation of the spin and charge gaps at weak but finite $U$
is then used to check the validity of the weak-coupling phase diagram.
The DMRG calculations of $U_c$ and the $U=\infty$ phase diagram were
reported in a preliminary form in Ref.\ \onlinecite{DaulNoack97}, but
here they are carried out with more accuracy and the discussion is
extended.


\section{THE \lowercase{$t_1-t_2$} CHAIN}
\label{modelprops}

\subsection{The model}

\begin{figure}[htb]
\begin{center}
 \epsfig{file=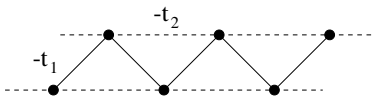,width=5cm}
\end{center}
\caption{The $t_1-t_2$ chain.}
\label{figt1t2chain}
\end{figure}

We consider the one-dimensional Hubbard model with
next-nearest-neighbor hopping (see Fig.\ \ref{figt1t2chain}) with
the Hamiltonian
\begin{eqnarray}
  \lefteqn{ H = -t_1 \sum_{i,\sigma} \left( c^{\dagger}_{i+1\sigma}
                                            c_{i\sigma} + h.c.  
 \right) } \nonumber \\
&&      -t_2\sum_{i,\sigma}\left( c^{\dagger}_{i+2\sigma} c_{i\sigma} 
                                  + h.c. \right)
    + U \sum_i n_{i\uparrow}n_{i\downarrow} . 
\label{eqhamt1t2}
\end{eqnarray}
We will call this model the  $t_1-t_2$  chain.

\begin{figure}[htb]
\begin{center}
 \epsfig{file=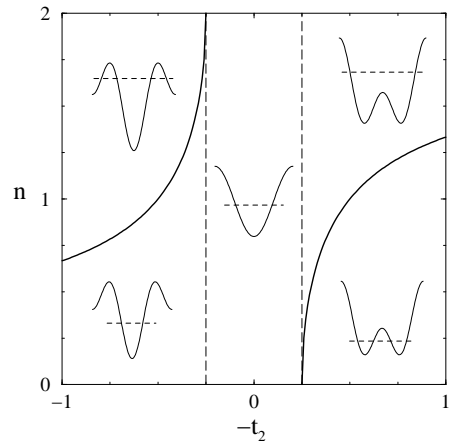,width=7cm}
\end{center}
\caption{$U=0$ phase diagram.
  The inset plots show the qualitative behavior of $\epsilon(k)$ in
  the different regions with the Fermi level indicated by the dashed
  line.
  } 
\label{figU0phasediag}
\end{figure}

The summation is over all $L$ sites and spin $\sigma$. 
Here we will always take $U$ positive and set $\hbar =1$.
The sign of $t_1$ is arbitrary since a gauge transformation 
$c_j \rightarrow e^{i\pi j}c_j$ can reverse it, so we set $t_1=1$
without loss of generality, and measure all energies in units of
$t_1$.
This Hamiltonian conserves the number of particles, the total spin
${\bf S}$ and its projection onto the quantization axis, $S_z$.
If a particle-hole transformation is applied to 
the system, the transformation $t_2 \rightarrow -t_2$ is necessary
to recover the original Hamiltonian. 
Therefore, the parameter regime $n > 1$, $t_2 > 0$ maps to  
$n < 1$, $t_2 < 0$ and the regime $n > 1$, $t_2 < 0$ maps to
$n < 1$, $t_2 > 0$ (see Fig.\ \ref{figU0phasediag}).
Because a definite order of the particles is no longer enforced when
$t_2\neq 0$, the Lieb-Mattis theorem \cite{Liebmattis62} does not
apply and, indeed, we will see that we do find ferromagnetism. 

The effect of the sign of $t_2$ can be understood by
considering the Hamiltonian of
Eq.\ (\ref{eqhamt1t2}) with $L=3$, $N=2$ particles, and open boundary
conditions.
This three-site model, treated by
Tasaki in Ref.\ \onlinecite{Tasaki96}, has a ferromagnetic ground state only
when $t_2<0$. 
For this choice of sign, the triangular structure frustrates the
antiferromagnetic order normally found in the Hubbard model.

For $U=0$ and periodic boundary conditions $H$ can be diagonalized by
Fourier transformation yielding
\begin{equation}
  H = \sum_{k,\sigma} \epsilon (k) c_{k\sigma}^{\dagger}c_{k\sigma}
\end{equation}
with $k$ an integer multiple of $\frac{2\pi}{L}$ and 
\begin{equation}
  \epsilon (k) = -2t_1\cos k -2t_2\cos 2k.
\end{equation}
The dispersion $\epsilon (k)$ will have one minimum at $k=0$ for 
$t_2 > -0.25$ and two minima at finite $k$ for $t_2 < -0.25$.
Similarly, there will be maxima in $\epsilon (k)$ at $k=\pm\pi$ for
$t_2 < 0.25$, but will shifted away from $k=\pi$ for $t_2 > 0.25$. 
Therefore, for small $|t_2|$, $\epsilon (k)$ does not differ
qualitatively from the $t_2=0$ band structure, and there are two
Fermi points for arbitrary electron density $n$.
On the other hand, for $t_2 < -0.25$ and sufficiently small densities
or for $t_2 > 0.25$ and sufficiently large densities, the Fermi 
surface has four Fermi points, namely $\pm k_{F_1}$ and $\pm k_{F_2}$
and, as we will see, can be mapped to a two-band model at weak
coupling.
The resulting $U=0$ ground-state phase diagram is depicted in 
Fig.\ \ref{figU0phasediag}.
Note that we take the horizontal axis to be $-t_2$ in this and all
subsequent phase diagrams in order to better display the $t_2 < 0$, 
$n < 1$ region on which we will primarily concentrate in this work.

\subsection{Square cluster}
\label{toysym}

\begin{figure}[htb]
\begin{center}
 \epsfig{file=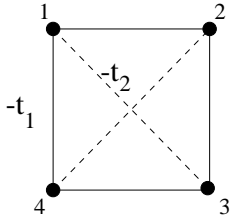,width=3cm}
\end{center}
\caption{Square cluster.}
\label{figsquare}
\end{figure}

It is useful to solve the Hamiltonian of Eq.\ (\ref{eqhamt1t2}) exactly
for $L=4$ and periodic boundary conditions in analogy with the solution of
the three-site problem discussed above.
This will allow us to further examine the effect of the sign of $t_2$
on the ground state, and also will illustrate finite-size effects due
to open and closed shell configurations.
As depicted in Fig. \ref{figsquare}, one obtains the Hubbard model on
a square with additional diagonal hopping.

We consider the $U=\infty$ limit with $N=2$ and $N=3$ electrons and minimum
$S_z$  (e.g. $S_z=0$ for $N=2$).
The dimension of the Hilbert space is 12 in both cases. 
With the aid of group theory, the problem can be solved analytically.
The Hamiltonian has the symmetry of the group $C_{4v}$ (see Table \ref{tablec4v}),
and can be diagonalized using the symmetry-adapted wave functions.
The eigenvalues are labeled by the corresponding irreducible
representations ($A_1, A_2, B_1, B_2, E$). 
In order to see whether a state is fully polarized
or not, we check whether the eigenfunction is symmetric or antisymmetric.
The irreducible representations $A_1, B_1$ and $B_2$ are symmetric and $A_2$
is antisymmetric. \cite{Atkins70}
Since the global wave function is required to be antisymmetric
and the spin function of a fully polarized state is symmetric, the 
ferromagnetic state must belong to the representation $A_2$.

An analysis of the eigenvalues (see Tables \ref{tableN2} and \ref{tableN3}) 
leads to the following conclusions:

\noindent $N=2 :$
For $t_2 > -0.5$  the ground state has $S=0$. In this case the fully polarized
state is not a closed shell state and it seems that for this reason it can not
be the ground state. 
Similar open shell effects have also been observed in numerical exact
diagonalization calculations for larger systems and other
fillings. \cite{DaulPieri97}

For $t_2 < -0.5$ (where the single-particle spectrum has two minima) the  
ferromagnetic ($S=1$) ground state of representation $A_2$ is degenerate 
with the non-magnetic ($S=0$) state of $B_1$. 
This is in agreement with Mielke's theorem, \cite{Mielke93} according to which
the ferromagnetic ground state is unique only when a restricted single-particle 
density matrix of the ferromagnetic  ground state is irreducible, and 
degenerate when it is reducible.
That the latter case applies for $L=4$ can be confirmed by explicit
calculation of this single-particle density matrix.
It is interesting to note that for $N=2$ on systems with $L>4$, it can
be shown numerically that the ground state is ferromagnetic and unique
when the single-particle spectrum has two minima.
This degeneracy therefore seems to be an artifact of the high
symmetry of the $L=4$ system.

\noindent
$N=3 :$
For any $t_2 < 0$ the ground state is ferromagnetic. This is actually the 
Nagaoka case of one hole in a half-filled band.

\subsection{Special limits for $U=\infty$}
\label{analyticallimits}

\begin{figure}[htb]
\begin{center}
 \epsfig{file=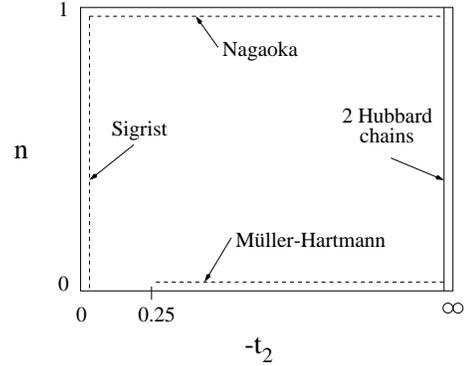,width=6cm}
\end{center}
\caption{Schematic $U=\infty$ phase diagram.}
\label{figdiagphaseinfty}
\end{figure}

For $U=\infty$ and negative $t_2$, ferromagnetism has analytically been shown 
to exist in three different limits.
For one hole in a half-filled band, the Nagaoka mechanism leads to a
ferromagnetic ground state; \cite{Mattis74}
for $|t_2| \rightarrow 0$, Sigrist and coworkers have shown 
that the model is ferromagnetic for all densities; \cite{Sigrist92}
and for $|t_2|>0.25$, where the band structure has two minima, 
M\"{u}ller-Hartmann \cite{MullerHartmann95} has shown that the
low-density limit is ferromagnetic.
These three limits are indicated in the schematic phase diagram 
shown in Fig. \ref{figdiagphaseinfty}. 
In addition, for $|t_2| \rightarrow \infty$ the model can be mapped onto
two decoupled Hubbard chains, which cannot be ferromagnetic due to the
Lieb--Mattis theorem. 

%


\section{Variational approach}
\label{variational}

The fully polarized state 
\begin{equation}
  | F\rangle = \prod_{k < k_F} c_{k\uparrow}^{\dagger} |0\rangle
\end{equation}
is an eigenstate of Hamiltonian (\ref{eqhamt1t2}) with energy
\begin{equation}
     E_{\mbox{\footnotesize{ferro}}} = \sum_{k<k_F} \epsilon_k , 
\end{equation}
where $\epsilon_k$ is the single-particle dispersion and $k_F$ is the
Fermi wave vector for the ferromagnetic state.  
The ferromagnetic state $|F\rangle$ is certainly unstable if a
variational state with one flipped spin and a lower energy can be
found. 
In order to put good constraints on the extent of a ferromagnetic
phase, it is important to use as good a variational wave function as
possible. A particularly sophisticated ansatz due to Edwards is defined by 
\begin{equation}
   \vert \chi \rangle  = \frac{1}{\sqrt{L}} \sum_{\ell=1}^{L} e^{iq\ell} 
      c_{\ell\downarrow}^{\dagger} \prod_{\alpha=1}^{N-1} 
         c_{\alpha\uparrow}^{\dagger}(\ell) \vert 0\rangle 
\label{eqedwardsAnsatz}
\end{equation}
where
\begin{equation}
  c_{\alpha\uparrow}^{\dagger}(\ell) =  \sum_{m=1}^{L} \varphi_{\alpha} (m-\ell) 
     c_{m\uparrow}^{\dagger} 
\end{equation}
creates an up-spin electron in an orbital that is determined variationally. 
The variational parameters are the wave vector $q$ and the $(N-1)L$
amplitudes $\varphi_{\alpha}(\ell)$. 
For $t_{2}=0$ this variational wave function includes the Bethe Ansatz as a 
special case, and is therefore an exact solution of the simple Hubbard 
chain.\cite{Edwards90}
This ansatz has previously been applied to the $t_1-t_2$ model in order to
calculate critical $U$ values in
Ref.\ \onlinecite{DaulPieri97}, but here we show the results of
additional calculations, including the full $U=\infty$ phase diagram.

The variational energy for orthonormal one-particle orbitals reads 
\cite{vonderLinden91}
\begin{eqnarray}
\lefteqn{ E(q, \{ \varphi_{\alpha}(\ell) \} ) =} \nonumber \\
 && - 2t_{1} \cos (q) \; \det  S^{(1)} -  2t_{2}\cos(2q)\; \det S^{(2)} 
    \nonumber \\
  && - 2t_{1} \; \mbox{tr} S^{(1)}  - \; 2t_{2} \;\mbox{tr} S^{(2)}\ 
   + \; U \sum_{\alpha ,\beta} \varphi^{*}_{\alpha}(0) \varphi_{\beta}(0)
\end{eqnarray}  
with overlap matrices $ S_{\alpha \beta}^{(\delta)} $ defined as
\begin{equation}
   S_{\alpha \beta}^{(\delta)} = \sum_{\ell=1}^{L} \varphi_{\alpha}^{*}(\ell)
         \varphi_{\beta}(\ell+\delta) .
\end{equation}
We use the conjugate gradient method \cite{Numrecipes} to minimize the 
energy. The derivatives 
\begin{equation}
 \frac{\partial E}{\partial\varphi_{\alpha}(\ell)}
\end{equation}
can be calculated analytically and are given in  Appendix \ref{gradient}. 
After each iteration the orbitals are orthonormalized
using the modified Gram-Schmidt method. 
The initial orbitals, which must be chosen so that they are
close enough to the final wave functions $\varphi_{\alpha}(\ell)$, are
taken to be the $N-1$ lowest eigenfunctions of the Hamiltonian (\ref{eqhamt1t2}) 
with $U=0$ and a site impurity at 0.
Care has to be taken in choosing the particle number $N$. 
In particular, the corresponding non-interacting ground state must be
non-degenerate (closed shell) in order to obtain
well-behaved convergence of the conjugate gradient calculation.

Here we want to determine the $U=\infty$ phase diagram. 
Since there is no double occupancy when $U=\infty$,
$\varphi_{\alpha}(0)=0$ for all $\alpha$ and the energy 
gradient with respect to $\varphi_{\alpha}(0)$ can be excluded. 
We also calculate the total spin  ${\bf S} = \sum_i {\bf S}_i$ of the
wave function.
Using the commutator $\left[ S^+,S^- \right] = 2S_z$ and working at a
particular $S_z$, we obtain
\begin{equation}
  \langle {\bf S}^2 \rangle = \frac{N}{2}\left( \frac{N}{2}-1\right)
  + \langle S^- S^+\rangle .
\label{eqs2s-s+}
\end{equation}
By applying Wick's theorem to the down spin operators, we can write
\begin{equation}
\langle S^-S^+ \rangle  = \frac{1}{L} \sum_{\ell,m} e^{iq(\ell-m)} \langle
  0| \prod_{\alpha} b_{\alpha}(\ell) \; c_{\ell\uparrow} c_{m\uparrow}^{\dagger}
\;   \prod_{\beta} b_{\beta}^+(m) |0\rangle .
\end{equation}
The elements of the sum can be expressed in term of a determinant
[cf. Eq. (A12) in Ref.\ \onlinecite{vonderLinden91}; care must be taken
since there are  typographical errors ], leading to
\begin{eqnarray}
  \langle S^-S^+ \rangle &=& 1 - \sum_{\delta=1}^{L-1} \cos(q\delta) 
  \det S^{(\delta)} \nonumber \\
  && \left[ \sum_{\alpha,\beta} \varphi_{\alpha}(\delta)
   {S_{\alpha\beta}^{(\delta)}}^{-1}  \varphi_{\alpha}(L-\delta) \right] .
\end{eqnarray}

\begin{figure}[htb]
\begin{center}
 \epsfig{file=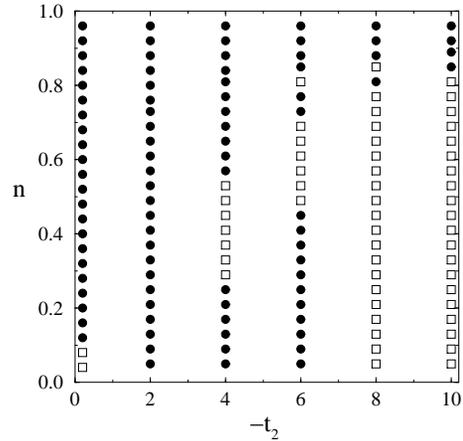,width=7cm}
\end{center}
\caption{$U=\infty$ phase diagram obtained with ansatz 
(\ref{eqedwardsAnsatz}) from numerical calculations on an $L=100$ lattice. 
The filled circles represent stable ferromagnetic 
states and the open squares  unstable ones. }
\label{figEdwardsphasediagram}
\end{figure}

For $S=S_{\mbox{\footnotesize{max}}}=N/2$,  (\ref{eqs2s-s+}) implies  
$\langle S^-S^+ \rangle =N$, whereas for $S=S_{\mbox{\footnotesize{max}}}-1$,
$\langle S^-S^+ \rangle $ must vanish.
When the ferromagnetic state is destabilized, we find that 
$\langle S^-S^+ \rangle \approx 0.01-0.1$.  
That this expectation value is not an eigenvalue of $S^2$
is an indication that the variational wave function is not an
eigenstate of $H$, as opposed to the $t_2 = 0$ case. 
The resulting $U=\infty$ phase diagram is displayed in
Fig. \ref{figEdwardsphasediagram}. 
As one can see from the figure, all four of the analytically
treatable limits of Sec.\ \ref{analyticallimits} are, for the most part,
reproduced, and the three ferromagnetic regions are connected to one
another.
One anomaly is that for small $|t_2|$ the ferromagnetic state is
destabilized at low density. 
As we will see, this does not occur in the DMRG calculations. 
This discrepancy is probably due to the large finite size effects on finite
lattices at low densities that come from the difference between closed
and open shells when the system has periodic boundary conditions.
We have already seen this behavior for $L=4$ in Sec.\ \ref{toysym}.
The alternating paramagnetic and fully polarized states as a function
of $n$ near the upper boundary of the paramagnetic phase are another
illustration of these finite-size effects.
In addition, for small $n$ and $|t_2| \ge 8$, the ferromagnetic 
state is unstable in a regime expected to be ferromagnetic by the
M\"uller--Hartmann argument. 
However, this argument is valid in the limit of small $n$, and here
the lowest obtainable $n$ is limited by $N/L$.


\section{EXACT DIAGONALIZATION}
\label{exactdiag}

We have performed numerical diagonalization using the Lanczos \cite{Lanczos}
and the Davidson \cite{Davidson} algorithms for chains of up to length $L=16$,
and various numbers of electrons $N$ and boundary conditions. 
These methods permit us to obtain both the numerically exact energy
and the wave function of the ground state on a finite cluster.

\subsection{Boundary conditions} 

On small lattices, it is important to analyze carefully the effect
of  boundary conditions. 
In order to understand which boundary 
conditions should be used, we consider the case of $N=2$ electrons on
a $L=12$ system with $t_2 = -0.1$. 
With periodic boundary conditions the system is never 
ferromagnetic, similar to the four-site model discussed in 
Sec.\ \ref{toysym}.
In fact, this seems to be the case for all lattice sizes $L$.
For anti-periodic boundary conditions, the single-particle spectrum
has 2 degenerate minima  at $k=\pm\frac{\pi}{12}$. 
Due to Mielke's theorem \cite{Mielke93} the model is
fully polarized for all $U>0$.
Therefore both boundary conditions are unable to reproduce the low-density 
regime with a paramagnetic ground state for small $U$ and a ferromagnetic 
ground state for large $U$.

Only with open boundary conditions do we obtain a non-magnetic ground
state for small $U$ and a ferromagnetic one for large $U$. 
We find that the critical value occurs at $U_c \approx 11$. 
Therefore, in order to minimize the effect of the boundary conditions,
we will take open boundary conditions
for all exact diagonalization calculations as well as for the DMRG
calculations described subsequently.

\subsection{Determination of $U_c$}
\label{DeterminationofUc}

In order to determine the critical value of $U$ above which the ferromagnetic
state is the ground state, we start at small $U$, for which the ground state is
not magnetic, and increase $U$ until we reach the fully polarized state.
The transition point can be determined by examining the behavior of
the energy $E_0$. 

\begin{figure}[htbp]
\begin{center}
 \epsfig{file=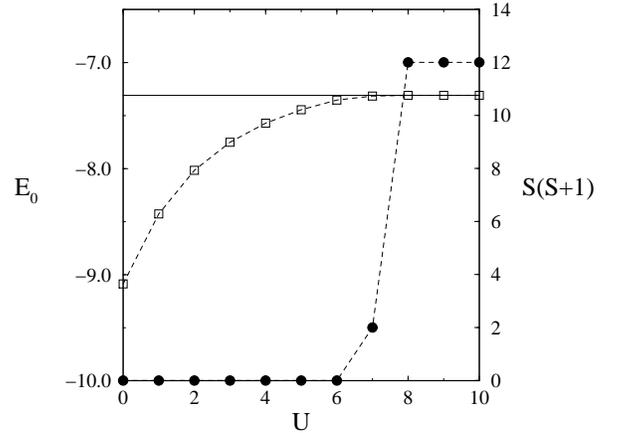,width=7cm}
\end{center}
\caption{$E_0$ (open squares) and $ \langle \psi_0 | {\bf S}^2 |\psi_0\rangle
=S(S+1)$ 
(filled circles) for a system of size $L=12$
with $N=6$ and $t_2=-0.2$. 
The horizontal solid line is the energy of the fully
polarized state. }
\label{figES2fU}
\end{figure}

As discussed in Sec.\ \ref{variational}, the energy of the fully polarized
ferromagnetic state, $E_{\mbox{\footnotesize{ferro}}}$, does not
depend on $U$ and is exactly known.
Thus, if  $E_0(U) = E_0(\tilde{U}) = E_{\mbox{\footnotesize{ferro}}}$ 
for all $U>\tilde{U}$, we identify $\tilde{U}$ with $U_c$.
This can be confirmed by verifying that the lowest eigenvalue is the same
in all $S_z$-subspaces, since then the ground state
must have a degeneracy of $2S_{\mbox{\footnotesize{max}}}+1$,
or by calculating the expectation value of the total spin operator in the 
ground state,
\begin{equation}
   \langle \psi_0 | {\bf S}^2 |\psi_0\rangle = \sum_{i,j}
   \langle \psi_0 | {\bf S}_i {\bf S}_j |\psi_0\rangle = S(S+1) .
\label{eqS2}
\end{equation}
For $U>U_c$ one will obtain $S= S_{\mbox{\footnotesize{max}}}$.
In Fig. \ref{figES2fU} we clearly see that the values for $U_c$ obtained using 
these two criteria are the same.

\subsection{Order of the transition}
\label{orderdavid}

In order to determine the order of the transition, we investigate with
very high precision the ground state energy $E_0(U)$ around
$U_c$. Since there are many states with energy very close to $E_0$ a
very large number of iterations are needed in the Davidson procedure
in order to obtain convergence 
(more than 1000 $H | \psi \rangle$ multiplications).

If the transition is first order, the ground state will jump from $S=0$
to $S=S_{\mbox{\footnotesize{max}}}$ and $E_0(U)$ will have a kink at $U_c$.
On the other hand, if the transition is second (or higher) order there
will be no kink in the energy as a function of $U$ and $S$ will smoothly
take on all values from 0 to $S_{\mbox{\footnotesize{max}}}$.
In the thermodynamic limit, a second order transition requires that
\begin{equation}
    \lim_{U\rightarrow U_c^-} \frac{\partial E_0}{\partial U} = 
   \lim_{U\rightarrow U_c^+} \frac{\partial E_0}{\partial U},
   \label{eqcontenergy}
\end{equation}
i.e. that the derivative of the ground state energy is continuous
through the transition.

\begin{figure}[htbp]
\begin{center}
 \epsfig{file=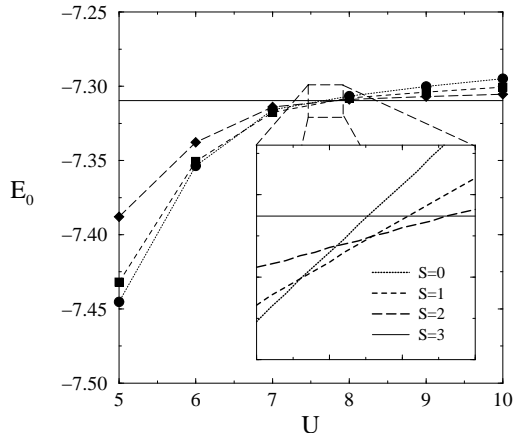,width=7cm}
\end{center}
\caption{$E_0(U)$ for $L=12$, $N=6$ and $t_2=-0.2$ for $S=0,1,2,3$.
The inset is a blowup of the indicated region.}
\label{figordertransition}
\end{figure}

In order to further clarify this issue we can follow the lowest energy
state with a particular spin $S$. 
Since utilizing the ${\bf S}^2$ quantum number in the exact
diagonalization program is technically difficult to implement, 
we follow a state of a particular ${\bf S}^2$ by diagonalizing the
augmented Hamiltonian 
\begin{equation}
   H' = H + \lambda {\bf S}^2
\end{equation} 
in different $S_z$-subspaces with $\lambda > 0$. 
For large enough $\lambda$, the lowest energy state within a given
$S_z$ sector will have the minimum $S$ value.
In Fig. \ref{figordertransition} we clearly see that the spin $S$ of
the ground state takes on all intermediate values as $U$ is increased.
This is an indication that Eq.\ (\ref{eqcontenergy}) will be satisfied
in the thermodynamic limit and that the transition is continuous.
Here we have chosen the parameters $n=0.5, t_2=-0.2$ so that the
system is in a regime with two Fermi points at $U=0$.
As will be shown in Sec.\ \ref{lowuphasediag}, the system is a
Luttinger liquid for weak $U$ at these parameters.
The transition is therefore from a Luttinger liquid to a ferromagnet.
Further evidence that the transition is second order based on the
behavior of the Luttinger-liquid parameters will be given in 
Sec.\ \ref{DMRG}.

\subsection{Small $t_2$}

\begin{figure}[htbp]
\begin{center}
 \epsfig{file=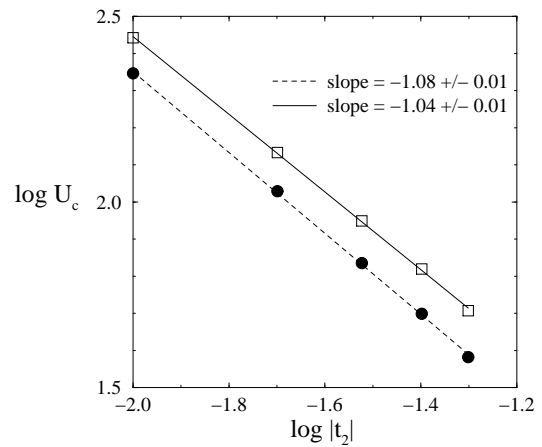,width=7cm}
\end{center}
\caption{$U_c$ as a function of $t_2$ for $L=10$ on a log-log scale. 
The filled circles are for $N=4$ and the open squares for $N=6$
while the solid and dashed lines are obtained using linear regression. }
\label{figuct2}
\end{figure}

We have already seen that for $t_2=0$ the fully polarized state always has a 
higher energy than the $S=0$ state unless $U=\infty$. 
For large but finite $U$, we can treat the model perturbatively in 
$\frac{1}{U}$. 
For $t_2=0$ and $n=1$ this yields the one-dimensional Heisenberg model. 
For the non-half-filled system and $|t_2| \ll t_1$,
we obtain, to first order in $\frac{1}{U}$,
\begin{eqnarray}
    H &=& -t_1\sum_{i,\sigma} \left( c^{\dagger}_{i+1\sigma}c_{i\sigma} 
   + h.c. \right)
  + \left( J_{\mbox{\footnotesize{eff}}}  + \frac{4t_1^2}{U} \right)  
   \sum_i {\bf S}_i  {\bf S}_{i+1} \nonumber \\
  && - \frac{t_1^2}{U} \sum_{i,\sigma,\sigma'} \lambda_{\sigma\sigma'}
   c^{\dagger}_{i+2\sigma}c^{\dagger}_{i+1-\sigma} c_{i+1-\sigma'}
  c_{i\sigma'} + h.c. .  
\label{eqtjham}
\end{eqnarray}
Here $\lambda_{\sigma\sigma} = +1$, $\lambda_{\sigma -\sigma} = -1$ and
\begin{equation}
 J_{\mbox{\footnotesize{eff}}} = \frac{t_2}{2\pi} \left[  
 \frac{2}{\pi n}\sin^2\pi n -    \sin 2\pi n  \right]
\end{equation}
is obtained as in Ref.\ \onlinecite{Sigrist92}. 
If $\sigma=\sigma'$ the third term of Eq.\ (\ref{eqtjham}) leads to a
permutation of the spin part of the wave function. 
The resulting matrix element (proportional to $\frac{1}{U}$) can be
incorporated into the second term to yield an effective
Heisenberg coupling
\begin{equation}
   \tilde{J} = J_{\mbox{\footnotesize{eff}}} + \frac{\gamma}{U}
\end{equation}
where $\gamma$ depends only on the filling of the system and, in
principle, can be calculated. 
Hence the coupling is ferromagnetic when
\begin{equation}
    -t_2 > \frac{\gamma}{U}
\end{equation}
and for small $|t_2|$ the critical $U$ should behave as
\begin{equation}
      U_c \sim |t_2|^{-1} .
\end{equation}
A numerical evaluation of $U_c$ in this low $|t_2|$ regime (shown in Fig. 
\ref{figuct2}) obtained from exact diagonalization is reproduced quite well by
this form.


\section{DENSITY MATRIX RENORMALIZATION GROUP}
\label{DMRG}

We now investigate much larger systems \cite{DaulNoack97} by applying the 
powerful and already widely used Density-Matrix Renormalization Group
technique. \cite{White92}
The DMRG is a variational procedure
which can be used to obtain the energies of the ground
state  and low-lying excited states very accurately, as well as to
compute a wide variety of equal-time correlation functions. 
Here we use the finite-size algorithm for system sizes of up to $L=140$
and keep up to 1000 states in the last iteration.
The efficiency of the algorithm is improved by keeping
track of the basis transformations in order to calculate a good
initial guess for the wave function
after adding a site to the system.\cite{White96}
Calculations applied to the $t_1-t_2$ model on small systems
show extremely good agreement with exact diagonalization results (up
to 10 figures).
The total discarded weight of the density-matrix eigenvalues provides
an estimate of the truncation error.
In the calculations performed here, the discarded density-matrix
weight ranges from $10^{-8}$ to $10^{-6}$.
The estimated error in the DMRG results shown here, determined by
examining the convergence with the number of states kept, is of the
order of the plotting symbol size or less, unless explicitly discussed.

We have also included the possibility of adding a term $\lambda {\bf S}^2$
to the Hamiltonian. 
Turning on $\lambda > 0$ shifts states of higher total
spin $S$ to higher energies.
This shift is known for a particular $S$ since ${\bf S}^2$ commutes
with the Hamiltonian.
This addition is useful for two reasons.
First, since the DMRG can only determine a limited number of excited
states accurately for a given number of states kept, it allows more
excited states to be accessed within a particular $S$ subspace. 
Second, it allows one to follow states of a particular $S$
individually, even if they are not the ground state of a particular
$S_z$ subspace for $\lambda=0$.
This trick is particularly useful near the 
ferromagnetic transition, where the ground state often has nonzero total
spin, and states with different total spin
are very close in energy.
The numerical problem of the mixing of different total $S$
states when they are near-degenerate in energy is also sometimes
relieved when $\lambda$ is turned on.

\subsection{$U_c$ as a function of the density}
\label{Ucdensity}

\begin{figure}[hbt]
\begin{center}
 \epsfig{file=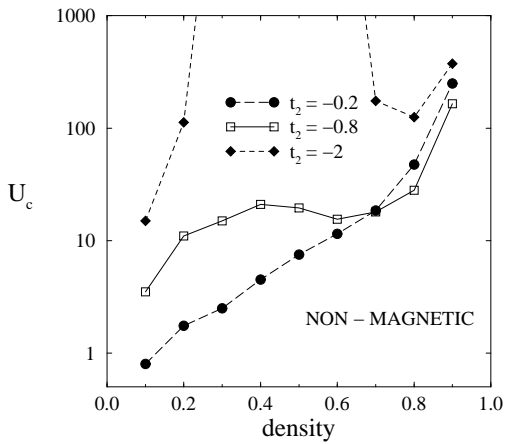,width=7cm}
\end{center}
\caption{$U_c$ as a function of density for $L=50$ on a log-linear scale for 
three different values of $t_2$.}
\label{figucn}
\end{figure}

\begin{figure}[hbt]
\begin{center}
  \epsfig{file=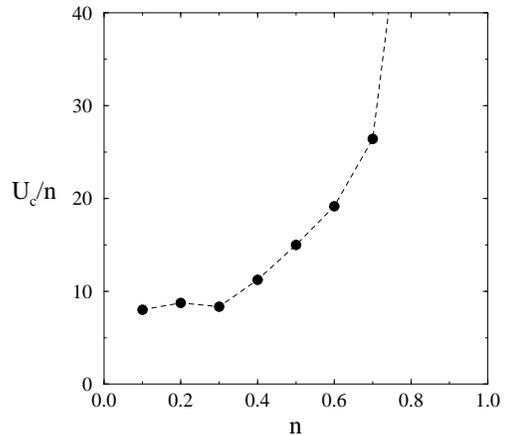,width=7cm}
\end{center}
\caption{Renormalized $U_c$ as a function of density for $L=50$ and $t_2=-0.2$.}
\label{figuceff}
\end{figure}

In order to determine $U_c$ we investigate the behavior of the ground
state energy and the expectation value of the total spin operator
$\langle {\bf S}^2 \rangle$ of the ground state, as in the
previously described exact
diagonalization calculations (see Sec.\ \ref{DeterminationofUc}). 
Here we have chosen to carry out the calculations on an $L=50$ lattice
because the finite-size effects are negligible upon further increasing
$L$.
In Fig. \ref{figucn} we
show results for various values of $t_2$ as a function of the density $n$.
The three curves show three different representative behaviors. 
In all cases, as $n\rightarrow 1$, $U_c$ diverges to reach the particular point
of $U=\infty$ where all states with different spin are degenerate.
For $t_2=-0.2$ (and all other cases with one minimum in the single-particle 
spectrum) $U_c$ increases monotonically with $n$. 
As $n\rightarrow 0$, $U_c \rightarrow 0$, which seems to imply that the
problem could be treated perturbatively in this limit.
However, the relevant parameter here is actually $U_c$ divided by $n$ which
tends to a finite value rather than going to zero at small densities
(see Fig. \ref{figuceff}).

For the two cases $t_2=-0.8$ and $t_2=-2$, a local minimum appears near
$n=0.6$ and $n=0.8$, respectively. This minimum in $U_c$ is due to 
a diverging density of states at the Fermi energy of the fully polarized state.
This occurs at a critical density $n_c$ where the Fermi energy coincides with
the local maximum in $\epsilon (k)$ at $k=0$, which exists only for 
$|t_2| >0.25$. For $t_2 =-2$ there exists a finite range of densities
in which the system never becomes ferromagnetic, even at $U=\infty$. 
Nevertheless, around
the critical density there still exists a finite $U_c$ with a minimum value
at $n_c$.

\subsection{Luttinger liquid parameters}
\label{Luttliqparam}

The Luttinger liquid concept \cite{Haldane81} 
(for a review see Ref.\ \onlinecite{Voit94} or
Ref.\ \onlinecite{Schulz91}) is based on a single-particle spectrum
with two Fermi points ($\pm k_F$).  
This is the case in the $t_1-t_2$ model for all densities when 
$|t_2| < 0.25$ and for large enough fillings when $t_2 < 0.25$ (see
Fig.\ \ref{figU0phasediag}).
The spectrum is linearized in the region of these two points and therefore
completely specified by the Fermi velocity $v_F$.

The interaction between electrons can then be written in terms of four
scattering processes: backward scattering ($g_1$), forward scattering
($g_2$), Umklapp scattering ($g_3$) connecting the region around
$+k_F$ with that around $-k_F$, and a $g_4$ term connecting states on
the same branch, either around $+k_F$ or around $-k_F$. 
The $g_4$ term, which is usually neglected since it leads only
to a renormalization of the Fermi velocity, will be seen to be
important here. 

We consider a non-half-filled system, where Umklapp process can be neglected.
Using bosonization, the Hamiltonian can be written in terms of boson
field operators $\phi_{\nu}$ and their canonically conjugate fields
$\Pi_{\nu}$ ($\nu=\rho,\sigma$ for charge and spin). 
This leads to the spin-charge separated Hamiltonian
\begin{equation}
   H = H_{\rho} + H_{\sigma} + \frac{2g_1}{\left( 2\pi\alpha \right)^2}
   \int dx \cos \left( \sqrt{8} \phi_{\sigma} \right) .
\end{equation}
Here $\alpha$ is a short-distance cutoff of the order of the lattice spacing and
\begin{equation}
  H_{\nu} = \int dx \left[ \frac{\pi u_{\nu}K_{\nu}}{2} \Pi_{\nu}^2 
 + \frac{u_{\nu}}{2\pi K_{\nu}} \left( \partial_x \phi_{\nu}  \right)^2 \right]
\end{equation}
is the Hamiltonian of an elastic string with eigenmodes corresponding to the
collective density fluctuations of the fermion liquid. 
The new parameters are the charge and spin velocities given by
\cite{Schulz91}
\begin{eqnarray}
   u_\sigma &=& \sqrt{{v_F}^2 - \left(\frac{g_1}{2\pi}\right)^2}  \\
  u_\rho &=& \sqrt{\left({v_F}+\frac{g_4}{\pi}\right)^2 
     - \left(\frac{g_1-2g_2}{2\pi}\right)^2} \label{eqvelocities}
\end{eqnarray}
and the two coefficients $K_{\rho}$ and $K_{\sigma}$ which determine
the asymptotic behavior of correlation functions
($K_\sigma=1$ for the Hubbard model due to the SU(2) spin symmetry). 

We can then calculate physical quantities such as the spin susceptibility
\begin{equation}
    \chi = \frac{2}{\pi} \frac{K_{\sigma}}{u_{\sigma}}
\label{eqchi}
\end{equation}
or the density-density correlation function 
\begin{equation}
\langle \delta n(x) \delta n(0) \rangle = -\frac{K_{\rho}}{(\pi x)^2}
+ A_1 \cos (2k_Fx) \frac{1}{x^{1+K_{\rho}} \log^{3/2}x}
+ \cdots , 
\label{eqdensdens}
\end{equation}
where $\delta n(x) = n(x) - \langle n(x) \rangle$.

\subsubsection{Determination of $K_\rho$}
\label{DeterminationKrho}

To obtain $K_{\rho}$ using DMRG we compute the Fourier transform 
\begin{equation}
  C(q) = \frac{1}{\sqrt{L}}
  \sum_{\ell} e^{iq\ell} N_{\mbox{\footnotesize{ave}}}(\ell)
\end{equation}
of the charge-charge correlation function (\ref{eqdensdens}) and then 
take the first derivative at $q=0$
\begin{equation}
\left . \frac{\partial C(q)}{\partial q} \right|_{q=0}
= \frac{K_{\rho}}{\pi} .
\label{eqkrho}
\end{equation}
The $q=0$ derivative is proportional to the coefficient of
the $1/x^2$ term in Eq.\ (\ref{eqdensdens}).
This method of extracting $K_{\rho}$ has been shown to yield accurate results
for the exactly solvable case $t_2=0$ and should be valid as long as
the system is in the Luttinger liquid phase. \cite{Dzierzawa95}
Since we work with open boundary conditions we have to use zero padding
for $|\ell| > L$
and average the correlation function to reduce boundary
effects \cite{Noack96}:
\begin{equation}
N_{\mbox{\footnotesize{ave}}}(\ell)
  = \frac{1}{n_a} \sum_{m=0}^{n_a-1} N (i_0+m,i_0+\ell+m) 
\end{equation}
where $i_0 + \ell/2= L/2$ and
\begin{equation}
   N(i,j) = \langle n_i n_j\rangle -  \langle n_i\rangle \langle n_j\rangle .
\end{equation}
The quantity $n_a$ is taken to be large enough so that 
$N_{\mbox{\footnotesize{ave}}}(\ell)$ does
not depend strongly on $i_0$ and $n_a$; typically we take $n_a\approx 6$.

\begin{figure}[hbt]
\begin{center}
  \epsfig{file=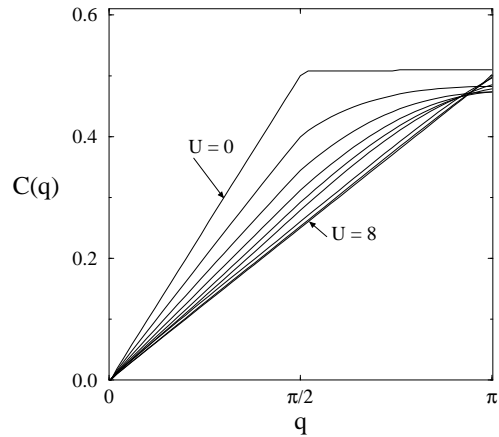,width=7cm}
\end{center}
\caption{Fourier transform $C(q)$ of the density-density correlation function
for systems of size $L=100$, $t_2=-0.2$, $n=0.5$ and $U=0,1, \ldots,8$.}
\label{figNq}
\end{figure}

\begin{figure}[hbt]
\begin{center}
  \epsfig{file=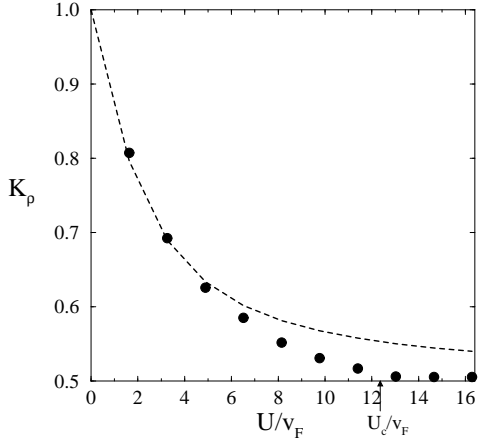,width=7cm}
\end{center}
\caption{$K_{\rho}$ as function of $U$ for $n=0.5$. The filled circles
  are for $t_2=-0.2$ and the dashed line is for $t_2=0$.}
\label{figKrho}
\end{figure}

The correlation functions for a system of size $L=100$ and $t_2 =-0.2$
with density $n=0.5$ are given in Fig. \ref{figNq}. 
One can see that the slope at $q=0$ is 
well-behaved and decreases monotonically with increasing $U$.
Therefore $K_{\rho}$ can be accurately calculated using
Eq.\ (\ref{eqkrho}). 
In Fig. \ref{figKrho} the numerical results for $t_2=-0.2$ are
compared with the well-known values for the simple ($t_2=0$) Hubbard chain,
\cite{Frahm90} where the appropriate parameter in the weak-coupling regime
is $U/v_F$. 
It is seen that the two curves agree very well for weak
couplings, although the Fermi velocities differ appreciably: 
$v_F(t_2=0)/v_F(t_2=-0.2) \approx 2.3$.
The deviation between the two curves increases for larger couplings.
For $t_2=0$, $K_\rho$ goes asymptotically to the value $\frac{1}{2}$
as $U \rightarrow \infty$, whereas for $t_2=-0.2$ $K_{\rho}$ reaches
$\frac{1}{2}$ at a finite $U$ whose value agrees quite well with the
$U_c$ calculated above.

\subsubsection{Spin and charge velocities}

To determine the velocity of the spin and charge excitations, we use
finite-size scaling of the corresponding energy gap
\begin{equation}
    \Delta_{\nu} = u_{\nu} \Delta k = u_{\nu}\frac{\pi}{L+1}
\end{equation}
where $\Delta k $ is the finite interval  between two adjacent $k$-points
for open boundary conditions.
The spin gap is defined as \cite{spinnote}
\begin{equation}
  \Delta_{\sigma} = E_0 (S=1) - E_0(S=0)
  \label{eqspingapdef}
\end{equation}
where $E_0(S)$ is the lowest eigenvalue for a system with spin S, and the
charge gap is given by
\begin{equation}
  \Delta_{\rho} = \frac{1}{2}\left[  E_0(N+2) + E_0(N-2) -2E_0(N) \right]
\end{equation}
where $E_0(N)$ is the ground state energy for $N$ particles and $S_z=0$. 

\begin{figure}[hbt]
\begin{center}
  \epsfig{file=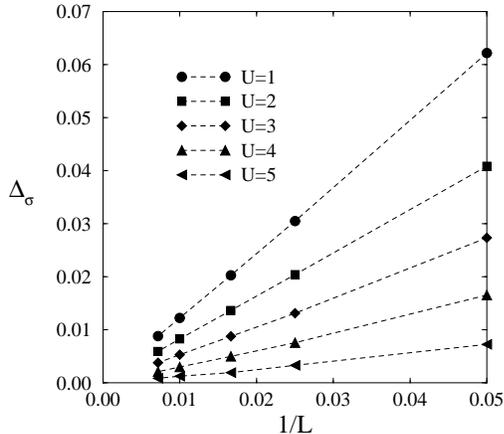,width=7cm}
\end{center}
\caption{Spin gap for different values of $U$ for systems with $t_2=-0.2$ and
$n=0.5$.}
\label{figspingap}
\end{figure}

\begin{figure}[hbt]
\begin{center}
  \epsfig{file=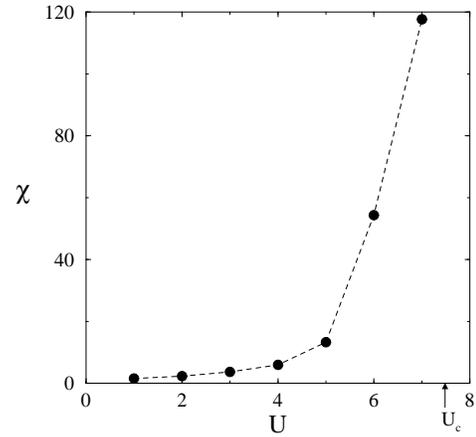,width=7cm}
\end{center}
\caption{Spin susceptibility as a function of $U$ for a system with $t_2=-0.2$ 
and $n=0.5$.}
\label{figchi}
\end{figure}

In Fig. \ref{figspingap}, we show the finite-size scaling of the spin gap
for a system with density $n=0.5$ and $t_2=-0.2$ for increasing values of $U$
(from 1 to 5). 
For these parameters, $U_c \approx 7.55$.
The excitations are gapless, as expected, and the spin velocity
$u_\sigma$ decreases to 0 with increasing $U$.
Here $K_{\sigma}=1$ due to the SU(2) invariance, and
the susceptibility (\ref{eqchi}) is then directly 
proportional to  $u_{\sigma}^{-1}$. 
Hence, as can be seen in Fig.\ \ref{figchi}, $\chi$ diverges when
approaching $U_c$.
A diverging susceptibility is an indication of a second-order
transition, in accordance with the analysis of Sec.\ \ref{orderdavid}. 
The critical exponent $\gamma$ is defined near the transition by
\begin{equation}
    \chi \sim u^{-\gamma},
\end{equation}
where $u=|U-U_c|$. 
We obtain $\gamma = 2.0 \pm 0.1$ by fitting the results
shown in Fig. \ref{figchi}, where the error is from the
least--squares fit.

\begin{figure}[htb]
\begin{center}
  \epsfig{file=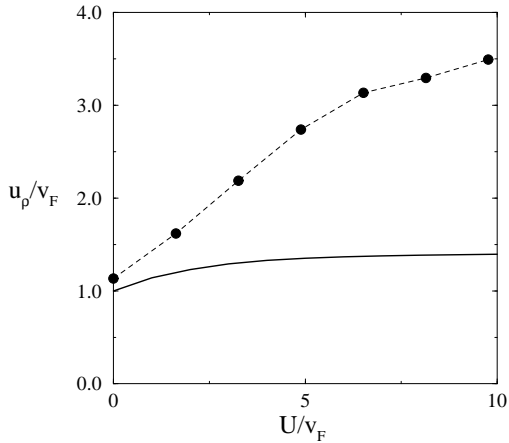,width=7cm}
\end{center}
\caption{The charge velocity $u_{\rho}$ as a function of $U$ for
  $t_2=0$ (full line) and  $t_2=-0.2$ (filled circles). }
  \label{figvelocities}
\end{figure}

We have also calculated the charge velocity. 
In Fig. \ref{figvelocities} we 
show the results for the same parameters as above and compare them
with the case $t_2=0$ (obtained following Ref.\ \onlinecite{Frahm90}).
We notice that for $t_2=-0.2$ the charge velocity, $u_{\rho}$, is
strongly renormalized from the $t_2=0$ value as $U$ is increased, even
when $U$ is rescaled by the Fermi velocity.
Eq.(\ref{eqvelocities}) suggests that this behavior could be governed
by the $g_4$ term which describes scattering between states on the
same branch of the spectrum and whose effect is usually taken to be
unimportant.
The increased importance of the $g_4$ interaction in a system with a
tendency towards ferromagnetism is understandable since the system's
response to an external magnetic field is described in terms of the
operator 
\begin{equation}
   O = \sum_{k,\sigma} \sigma \left[ c_{(k_F+k)\sigma}^{\dagger}
       c^{\phantom{\dagger}}_{(k_F+k)\sigma} +
      c_{(-k_F+k)\sigma}^{\dagger}
       c^{\phantom{\dagger}}_{(-k_F+k)\sigma}    \right]
\end{equation}
which involves states on the same branch of the spectrum. 
This indicates that $g_4$ must  
appear in the renormalization of the corresponding response function.
Hence the $g_4$ term must be relevant for the ferromagnetic fixed
point, and we would therefore expect $u_\sigma$ to be strongly
rescaled.

\subsection{Spin-spin correlation function}

\begin{figure}[htb]
\begin{center}
  \epsfig{file=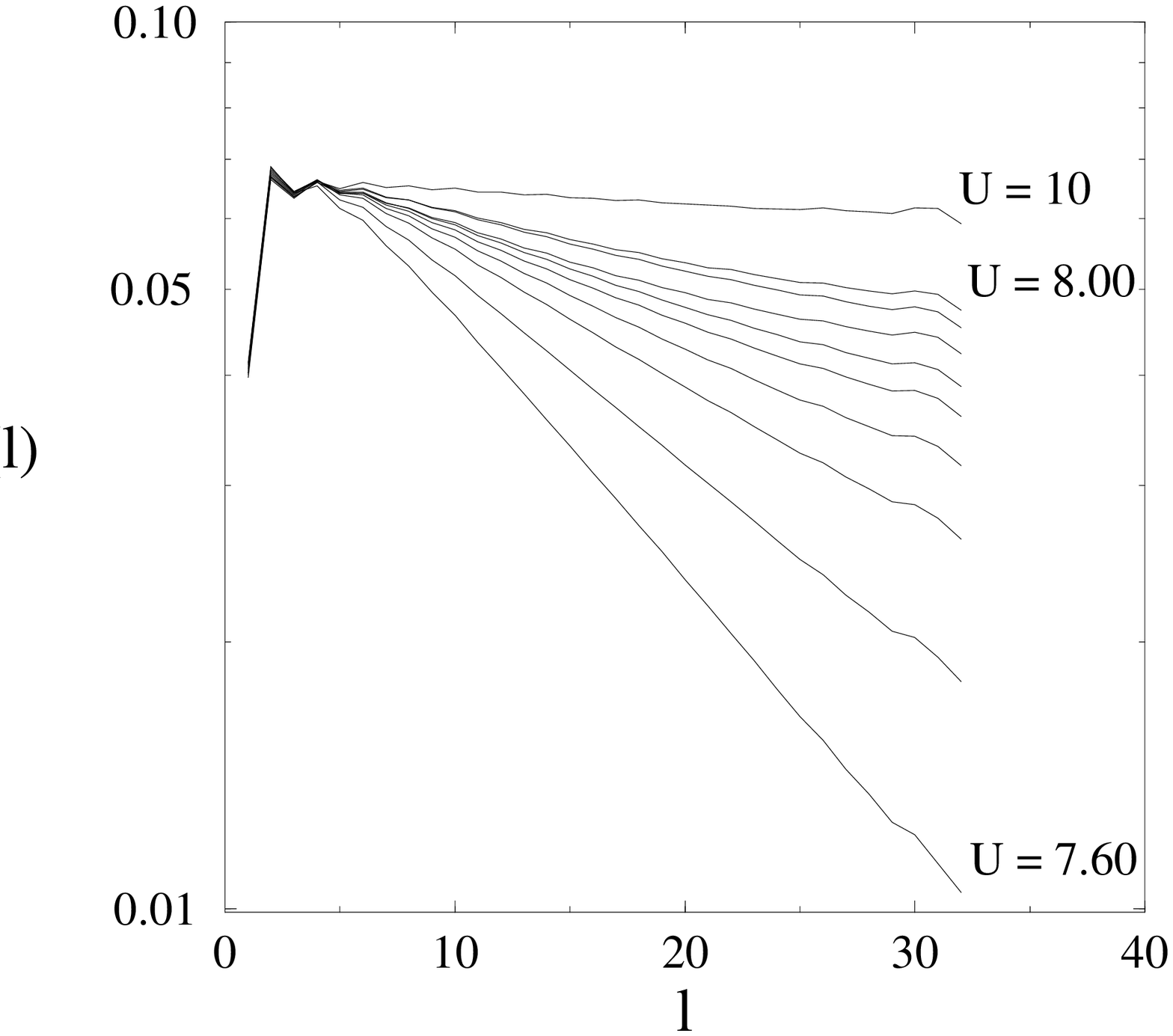,width=7cm}
\end{center}
\caption{The averaged spin-spin correlation function  
 $S_{\mbox{\footnotesize{ave}}}(\ell)$ for a system with $L=40$, $t_2=-0.2$ and
$n=0.5$ for $U=7.60,7.65,\ldots,8.00$ and 10, plotted on a semilog scale.}
\label{figsij}
\end{figure}

We can also study the behavior of the spin-spin correlation function near the 
transition. 
In order to minimize effects of the open boundary conditions, we
average over a number of sites for a given distance, as was done for
the density-density correlation function in
Sec. \ref{DeterminationKrho}, and consider
\begin{equation}
S_{\mbox{\footnotesize{ave}}}(\ell)
  = \frac{1}{n_a} \sum_{m=0}^{n_a-1} S (i_0+m,i_0+\ell+m) ,  
\end{equation}
where $i_0 + \ell/2= L/2$, $n_a \approx 4$ and
\begin{equation}
   S(i,j) = \langle S_i^-S_j^+ \rangle .
\end{equation}
Fig. \ref{figsij} shows the result for values of $U$ near $U_c$.
We find that $S_{\mbox{\footnotesize{ave}}}(\ell)$ is positive definite,
indicating ferromagnetic correlations, and can be well-fitted
by the form $e^{-\ell /\xi}$, with $\xi$ diverging as the transition is
approached.
This is seen by the linear behavior of $S_{\mbox{\footnotesize{ave}}}(\ell)$
on the semi-log plot, with decreasing slope as $U$ increases and the transition
is approached.
We can define the critical exponent associated with the divergence of the
correlation length as
\begin{equation}
    \xi \sim u^{-\nu}
\label{eqnu}
\end{equation}
where $u = |U-U_c|$.
Unfortunately, near the transition mixing of energetically close
states make it numerically difficult to accurately calculate 
$S_{\mbox{\footnotesize{ave}}}(\ell)$, even using 7 iterations and
keeping up to 800 states.
There tend to be systematic errors, which we minimize by
limiting the system size to $L=40$.
However, $\xi$ then quickly becomes of the order of the system size as
the transition is approached,
so that we can at most say that the best fit to 
the data occurs with a relatively small value of $\nu$, $\nu \approx 0.2$.

The total spin $S$ of the ground state is also related to this
correlation function (see Eq.\ (\ref{eqS2})) via
\begin{equation}
  \langle {\bf S}^2 \rangle = S(S+1) = S_z(S_z+1)
  + \sum_{i,j} \langle S^-_i  S^+_j \rangle , 
\end{equation}
which for a translationally invariant system becomes
\begin{equation}
  S(S+1) = S_z(S_z+1) + L \sum_{\ell} S(\ell) ,
\end{equation}
where $S(\ell) = \langle S^-_i  S^+_{i+\ell} \rangle$ is independent
of $i$.
Therefore, the ferromagnetic order parameter is
\begin{equation}
  s = S/L \sim L^{-1/2} \left [ \sum_{\ell} S(\ell) \right ] ^{1/2}.
\end{equation}
If $\sum_{\ell} S(\ell)$ is finite for $L \rightarrow \infty$, the system is 
disordered; 
if it is proportional to $L$, there is long-range ferromagnetic order; and if
it follows a power law $L^{2-\eta}$ with $1 < \eta < 2$, the system
is at the critical point with critical exponent $\eta$ and
$S(\ell) \sim \ell^{1-\eta}$.
In the ordered ferromagnetic phase, we do find a nonzero value of
$S(\ell)$ at large distances as seen in Fig.\ \ref{figsij}, consistent with
this picture, but we 
have not been able to determine the critical exponent $\eta$ from
numerical calculations because of the poor convergence of the DMRG at
the critical point and because of uncertainty in $U_c$.


\section{PHASE DIAGRAM}
\label{phasediagrams}

\subsection{$U=\infty$ phase diagram}
\label{infphasediag}

The phase diagram for $U=\infty$ can be determined using DMRG calculations. 
While one could, in principle, start with a Hilbert space in which
double occupancy is explicitly excluded, here we simply set $U=10^6$
to mimic the infinite-$U$ limit and find that the accuracy is
quite good since the DMRG integrates out high 
energy scales automatically. 
In order to decide if a point in the $n-t_2$ plane is 
ferromagnetic or not, we compare its DMRG energy $E_D$ with the known
ferromagnetic 
energy $E_{\mbox{\footnotesize{ferro}}}$ and, in addition, calculate $\langle 
{\bf S}^2  \rangle = S(S+1)$ and compare it with  
$S_{\mbox{\footnotesize{max}}}$. 
If $S \approx S_{\mbox{\footnotesize{max}}}$  
and $E_D > E_{\mbox{\footnotesize{ferro}}}$ for given $t_2$ and $n$ we 
conclude that the system is ferromagnetic for these parameters, whereas if we 
find an energy $E_D$ lower than $E_{\mbox{\footnotesize{ferro}}}$ and 
$S\approx 0$, we conclude that it is non-magnetic.  
Near the boundary between these two regions 
we sometimes find partly polarized states, i.e. 
$E_D < E_{\mbox{\footnotesize{ferro}}}$ but $S>0$. 
It is difficult to determine the nature of these states because there are
two possible causes for the partially polarized value of $S$.
One possibility is that the system undergoes a continuous phase
transition as a function of $n$ or $t_2$ at $U=\infty$.
However, near a ferromagnetic phase transition,
near-degeneracy of states leads to a mixing of states in the
diagonalization step of the DMRG procedure.
Therefore, the presence of partially polarized states could also be due to
numerical effects.
This mixing can also be seen in that the values of $S$
obtained in general take on continuous values that lie between
the discrete values of $S$ allowed by the finite number of electrons
in the system.
These problems are similar to those that occur near the ferromagnetic
transition at finite $U$ discussed in Sec.\ \ref{orderdavid}, where we
used extremely accurate exact diagonalization calculations instead of
DMRG to solve them.

\begin{figure}[hbt]
\begin{center}
  \epsfig{file=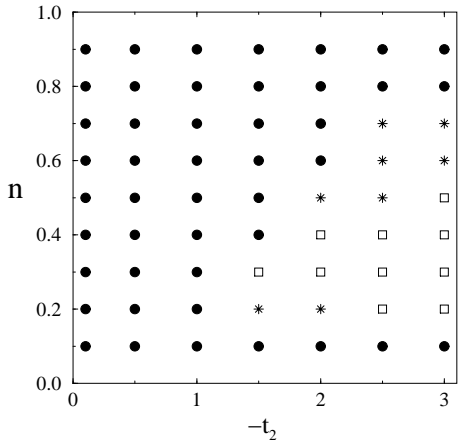,width=7cm}
\end{center}
\caption{Phase diagram for $U=\infty$ and $L=30$ obtained with DMRG. The filled circles
  indicate a fully polarized ground state, the open square a
  non-magnetic ground state and the stars a partially polarized ground state.}
\label{figdmrgdiagphase}
\end{figure}

As seen in Fig.\ \ref{figdmrgdiagphase},
we obtain a large region of ferromagnetism.
The three limiting cases of Sec.\ \ref{analyticallimits} are reproduced and
the corresponding ferromagnetic regions are, in fact, connected.
We also notice that for sufficiently large $|t_2|$, the system is not
magnetic at intermediate densities. 
This extends the limit $t_2\rightarrow-\infty$ in which
the system consists of two uncoupled Hubbard chains which must be non-magnetic 
due to the Lieb-Mattis theorem, \cite{Liebmattis62} to a finite region.
In this region, the system behaves effectively like an uncoupled
two-chain model.
We will later present evidence that this region in which the system
behaves as two uncoupled
Hubbard chains extends to the low-$U$ phase diagram.
Note that this phase diagram is qualitatively similar to that obtained
using the Edwards ansatz (see Fig.\ \ref{figEdwardsphasediagram}),
except that here the entire $t_2$ axis is
ferromagnetic for small densities. 

\subsection{Low-$U$ phase diagram}
\label{lowuphasediag}

\begin{figure}[hbt]
\begin{center}
  \epsfig{file=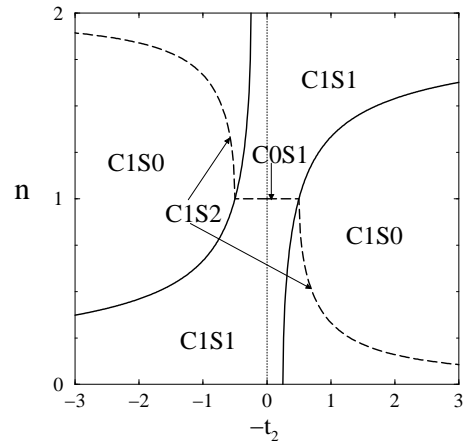,width=7cm}
\end{center}
\caption{Low-U phase diagram obtained by adapting the results of 
  Ref. 32. 
  Some small regions of additional phases
  near the solid lines are not shown.}
\label{figgendphaselowU}
\end{figure}

\begin{figure}[hbt]
\begin{center}
  \epsfig{file=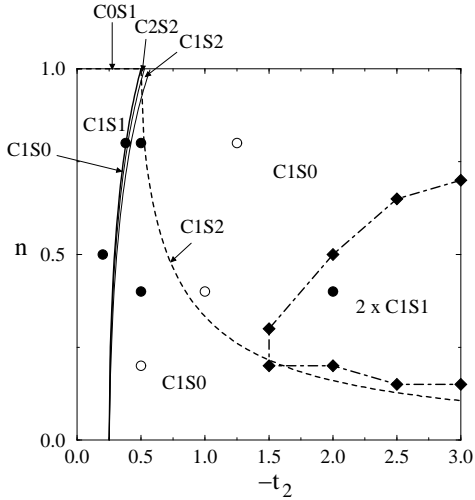,width=7cm}
\end{center}
\caption{Low-$U$ phase diagram restricted to negative $t_2$ and $n<1$.
The filled circles are for parameters for which no spin gap has been
found with DMRG at $U=2$, 
while the open circles are for parameters where a spin gap has been found.}
\label{figdphaselowU}
\end{figure}

We now turn to the general question of the low $U$ phase diagram within a
weak-coupling analysis. 
Balents and Fisher \cite{Balents96} have analyzed the
weak-coupling phase diagram of the two-chain Hubbard model using RG and 
bosonization. 
Their calculation is generic for a system with four Fermi points.
They obtain coupled RG equations and integrate them numerically 
to find the different fixed points, which they then analyze using bosonization.
The possible phases can be classified by the number of charge and spin modes
which are gapless. 
A phase with $\alpha$ gapless charge modes and
$\beta$ gapless spin modes is denoted C$\alpha$S$\beta$, where
$\alpha$ and $\beta$ can take on integer values from 0 to 2.
An adaption of their two-chain phase diagram to the $t_1$-$t_2$ model
in the $t_2$-$n$ plane
is shown in Fig.\ \ref{figgendphaselowU}, with
the $t_2 < 0$, $n < 1$ quadrant shown in more detail in 
Fig.\ \ref{figdphaselowU}. 

As discussed in Sec.\ \ref{Luttliqparam}, a Luttinger liquid phase
(C1S1) is expected for small $|t_2|$ in the region where the $U=0$
system has two Fermi points.
At half-filling $2 k_{F} = \pi$ and umklapp processes cause the
system to open a charge gap, so that the phase is that of the
one-dimensional Heisenberg model (C0S1). 

\begin{figure}[hbt]
\begin{center}
  \epsfig{file=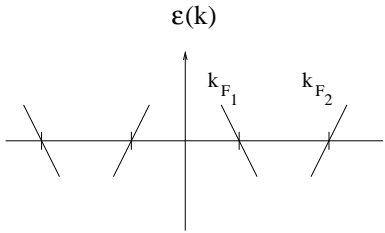,height=3cm}
\end{center}
\caption{Effective low-energy model for a Fermi surface with four points. }
\label{figeffmodel}
\end{figure}

When the Fermi surface has four points, namely $\pm k_{F_1}$ and
 $\pm k_{F_2}$, the effective low-energy model has four linearized
single-particle branches $\epsilon_1(k) = \mp v_{F_1} (k-k_{F_1})$ and
$\epsilon_2(k) = \pm v_{F_2} (k-k_{F_2})$ (see Fig. \ref{figeffmodel}).
This is equivalent to the two-band model of 
Ref.\ \onlinecite{Balents96} for which there are also four Fermi points, 
namely $\pm k_F^b$ for the bonding, and $\pm k_F^a$ for the
antibonding band. 
The only difference here is that the inner bands, which are denoted
$\epsilon_1(k)$ in the present case and originate from the antibonding
band in the two-chain model, have opposite velocities. 
The correct mapping of the Fermi points between the two models is therefore
\begin{eqnarray}
  \pm k_{F_1} & \rightarrow & \mp k_F^a \nonumber \\
  \pm k_{F_2} & \rightarrow & \pm k_F^b \nonumber .
\end{eqnarray}
Provided that this mapping is performed, the perturbation expansion
of the $t_1-t_2$ model and the two-chain Hubbard model is exactly the same at 
low energy, as already pointed out by Fabrizio. \cite{Fabrizio96} 
Therefore, we can simply adapt the results from 
Ref.\ \onlinecite{Balents96} to our case. 
In Fig.\ \ref{figgendphaselowU} and Fig.\ \ref{figdphaselowU},
the thick solid line represents the critical
density $n_c$ for which the Fermi surface splits into four points
(cf. Fig.\ \ref{figU0phasediag}). 
Exactly on this line, Balents and Fisher predict a C1S0 phase. 
For slightly smaller densities, $v_{F_1}$ 
is much smaller than $v_{F_2}$ leading  first to a C2S2, then to a
C1S2 phase.
(These three phases are not depicted in Fig.\ \ref{figgendphaselowU} since
they have a small extent.)
When $v_{F_1}$ is comparable to $v_{F_2}$ the weak-coupling RG leads
to a large region of a C1S0 phase.
This phase is a doped spin-liquid phase with a spin gap in which
power-law pairing and CDW correlations coexist. \cite{Dagotto97}
When $2 k_{F_2} = \pi$ within this region, indicated by a dashed line,
umklapp processes in the corresponding bonding band can open a charge
gap.
For the two-chain model, Balents and Fisher predict a 
C1S2 phase along some of this line, with the result being sensitive to
the initial conditions in the RG equations, i.e. on $v_F$ and
the initial couplings. 

In order to investigate the validity of this weak-coupling phase
diagram, we have calculated the spin and charge gaps using the DMRG
for different system sizes at small $U$ (we choose $U=2$) and a number of 
$t_2$ and $n$ values.
Due to the weak coupling and small size of the gaps, very high
precision is necessary in the DMRG procedure.
We use up to 8 finite-size iterations and keep up to 800 states in
the last iteration. 
The presence of a spin gap in the extrapolated $L\rightarrow\infty$
limit is indicated by an open circle in Fig. \ref{figdphaselowU} and
the absence of a spin gap by a solid circle.

\begin{figure}[hbt]
\begin{center}
  \epsfig{file=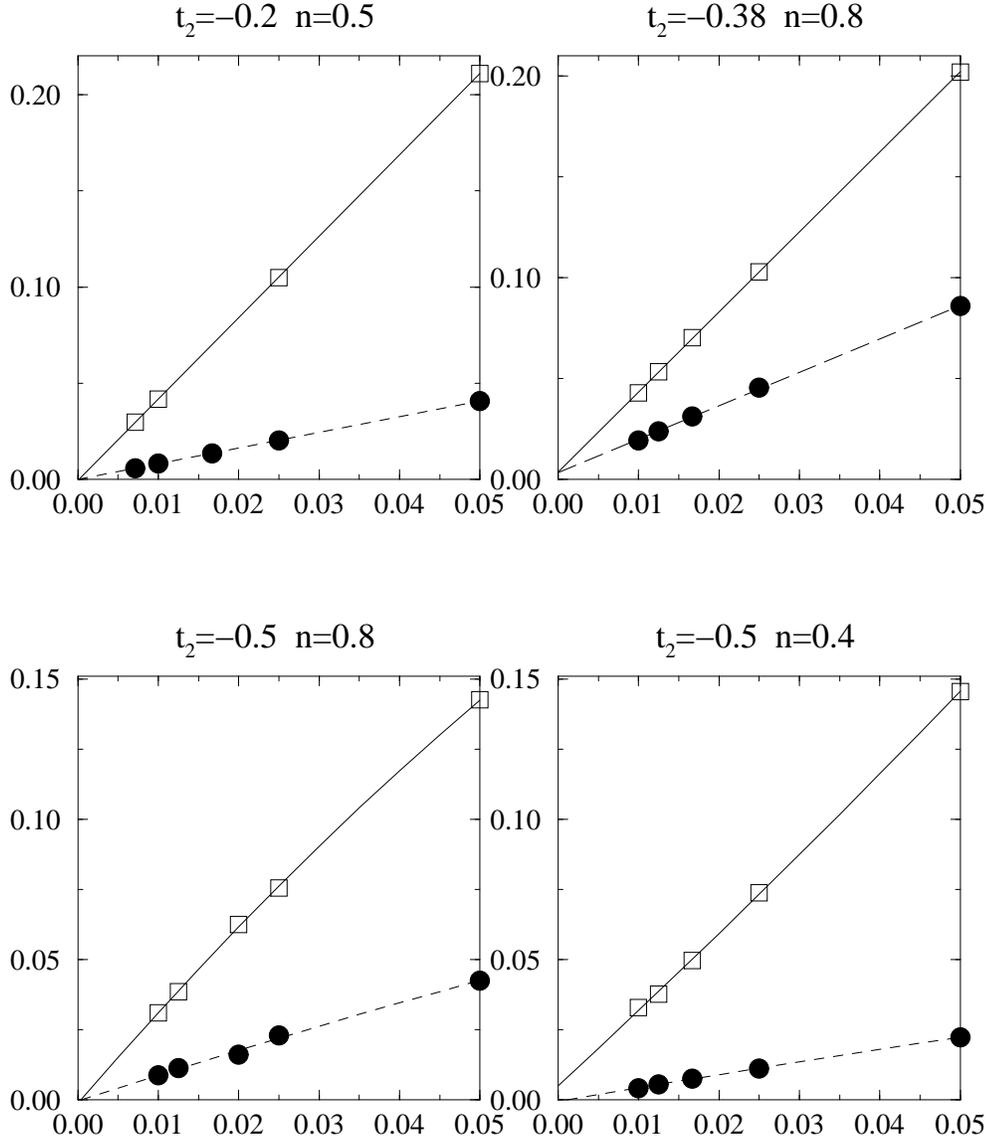,height=16cm}
\end{center}
\caption{Finite-size scaling of charge (open squares) and spin (filled
  circles)  gap for different parameters.}
\label{figgapscaling1}
\end{figure}

\begin{figure}[hbt]
\begin{center}
  \epsfig{file=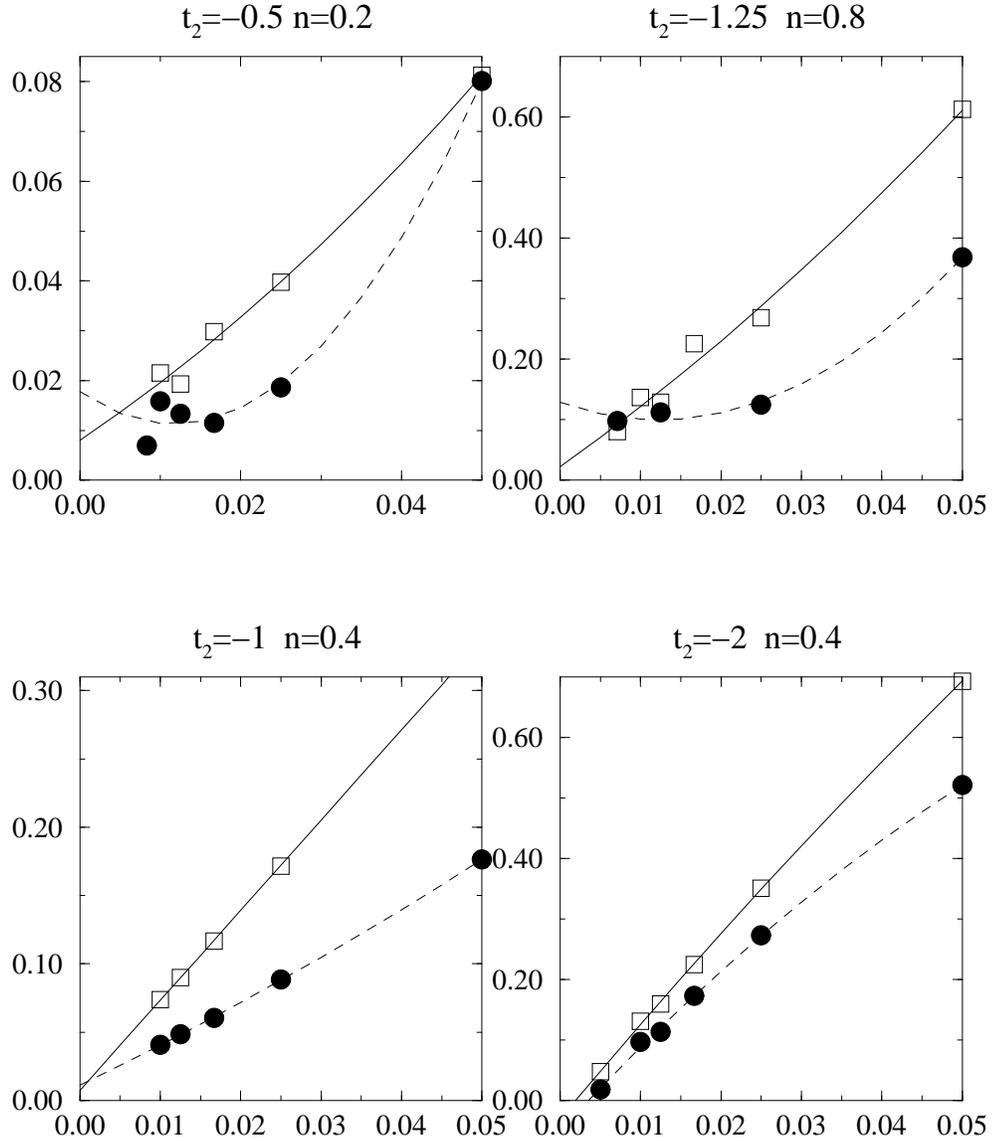,height=16cm}
\end{center}
\caption{Finite-size scaling of charge (open squares) and spin (filled
  circles) gap for different parameters.}
\label{figgapscaling2}
\end{figure}

Figs.\ \ref{figgapscaling1} and \ref{figgapscaling2} show the finite-size
scaling of the charge and spin gap.
The filled circles represent the spin gap and the open squares the
charge gap while the lines are quadratic regression in $(\frac{1}{L})$
between the points.
In general, the finite-size corrections to the spin and charge gaps
for a system with open boundaries can be represented as a power series in
$1/L$. 
When a gap is present, the dominant correction is usually $1/L^2$, and
when the system is gapless, the dominant correction\cite{White94} is $1/L$. 
While this behavior is generally seen in Figs.\ \ref{figgapscaling1}
and \ref{figgapscaling2}, there are cases with a small gap with an
obvious positive quadratic correction, but also a substantial
linear term.
In addition, there is scatter of up to the order of the symbol size in
some of the curves which we believe is due to additional finite-size
effects which can oscillate with the number of particles.

\onecolumn

For ${t_2=-0.2, n=0.5}$, the clearly vanishing charge and spin gaps
confirm that the system behaves as a Luttinger liquid
(see Sec.\ \ref{Luttliqparam}). For parameters close to the line of
critical $n$, namely ${t_2=-0.5, n=0.4}$ and ${t_2=-0.5, n=0.8}$,
we find no spin and charge gap, in contradiction  with the adapted
weak-coupling phase diagram.
However, a finite $U$ could renormalize the band structure, leading to
a shifting of the line of transition from two Fermi points to four.
Such a shift is seen with increasing $U$ in the two-chain Hubbard
model. \cite{Noack97}
Precisely on this transition line, for ${t_2=-0.38, n=0.8}$, a
spin gap would be predicted, but we do not find one.  
However, this phase might be hard to see numerically since it occurs
only exactly on the line, or might not be present for finite $U$.
Such a phase has also not been found numerically in the two-chain
Hubbard model. \cite{Noack96}

For the case ${t_2=-1.25, n=0.8}$ we find a spin gap, as predicted. 
For ${t_2=-0.5, n=0.2}$, where a spin gap is also predicted, we
have taken $U=0.5$ rather than $U=2$ because the ground state is
not paramagnetic at the larger $U$ value due to the proximity of the
ferromagnetic transition.
At the smaller $U$, finite size effects and numerical problems make it
difficult to definitely determine whether or not there is a spin gap.
Nevertheless, since there is a strong quadratic correction, we conclude
that a spin gap is probably present. 

Near the line $2k_{F_2}=\pi$, (i.e. at 
${t_2=-1, n=0.4}$) the system has a small spin gap, consistent with the
C1S0 phase but not a possible C1S2 phase.
However, as pointed out above, the existence of the C1S2 phase in the
weak--coupling calculation is dependent on the initial conditions, so
it may not be present along all of the dashed line.
In the two-chain Hubbard model, numerical DMRG calculations
\cite{Noack96} do find evidence for this phase for some fillings, even
at intermediate to strong $U$.
More work would have to be done to determine in detail some of the
finer structure of the phase diagram, but the difficulty of the
calculations and the finite-size scaling preclude a more detailed
investigation here.

For $t_2=-2.0$ and $n=0.4$, we find that both the spin and the charge
gaps vanish, in contradiction to the weak-coupling phase diagram,
which would predict a C1S0 phase.
This occurs in the regime which is paramagnetic at large $U$ because
the system behaves as two uncoupled chains (see Sec.\ \ref{infphasediag}).
We therefore suspect that the strong-coupling behavior extends to weak
coupling, and that the system is in a 2 $\times$ C1S1 = C2S2 phase here.
The phase boundary of the paramagnetic strong-coupling phase,
indicated by the solid diamonds, is also sketched in.
It remains to be determined how much this phase boundary changes in
going from strong to weak $U$.

\twocolumn

It should also be noted that Kuroki et al. \cite{Kuroki97} have studied the
$t_1$-$t_2$ model numerically using Projector Quantum Monte Carlo and
the DMRG at half-filling and $t_2 =-0.8$ and find
a spin-gapped metallic phase with dominant pairing correlations at
weak $U$, in agreement with Fig.\ \ref{figgendphaselowU} and 
Ref.\ \onlinecite{Fabrizio96}.
As $U$ is increased they find a transition to a spin-gapped insulating
phase also in agreement with Ref.\ \onlinecite{Fabrizio96}.
Here we have concentrated on $n < 1$ in the numerical work since the
ferromagnetic phase is not present at $n=1$.
In very recent work that we became aware of as this manuscript was being
completed, Arita et al. \cite{Arita97}, have calculated the spin gap at
$n=1.0$ and $n=0.5$ and $t_2=-0.55,-0.8,-2.0$ using the DMRG.
Of particular interest for the weak-coupling phase diagram are the
$n=0.5$, $t_2= -0.8,-2.0$ points which are in a paramagnetic phase.
For the $t_2= -0.8$ point, they find a very small but finite
gap at $U=8$, consistent with a C1S0 phase.
However, we believe that their data are also consistent with a
vanishing spin gap, which would be consistent with the C1S2 phase
present along the dashed line in Fig.\ \ref{figdphaselowU}.
At $t_2= -2.0$ and $U=16$, they find a vanishing spin gap, in
contradiction with the weak-coupling prediction of a gapped C1S0 phase
at these parameters, but consistent with the $2\times\mbox{C1S1}$ phase
proposed above.

\section{CONCLUSION}

We have studied an extended version of the conventional
one-dimensional Hubbard model in order to investigate the mechanism
for ferromagnetism in an itinerant electron model. 
The added term, which involves hopping between next-nearest neighbor
sites, precludes the application of the Lieb-Mattis
theorem \cite{Liebmattis62} which excludes a ferromagnetic ground
state in the one-dimensional Hubbard model. 
Indeed, we do find a ferromagnetic phase in a wide region of
parameters at large enough $U$ in the regime with $t_2$ negative and
$n < 1$ (which is equivalent to the $t_2 > 0$, $n > 1$ region through
a particle-hole transformation).

Using exact diagonalization, the Density-Matrix Renormalization Group and the 
Edwards variational ansatz, we have shown
that the three different mechanisms for ferromagnetism obtained by
taking special limits at $U=\infty$ (the Nagaoka state, the limit of
vanishing density and the limit of very small $t_2$) are all connected
in the same phase in the $t_2$--$n$ plane.
For large negative $t_2/t_1$, there is a paramagnetic region in the
large--$U$ phase diagram in which the system behaves like two
independent Hubbard chains.
This region extends to $t_2 \approx -1.5$ at intermediate $n$.

The critical interaction strength for the ferromagnetic transition,
$U_c$, has three characteristic behaviors as a function of $n$.
When $0 > t_2 > -0.25$, there is one minimum in the single-particle
dispersion, $\epsilon(k)$, and $U_c$ increases monotonically with $n$.
For $-0.25 > |t_2| > -1.5$, there are two minima in $\epsilon(k)$, and
there is a local minimum in $U_c$ when the Fermi level of the
fully polarized ferromagnetic state is at the
singularity in the density of states corresponding to the local
maximum in $\epsilon(k)$.
Finally, when $|t_2| < -1.5$, there is an intermediate region on $n$
for which there is no ferromagnetism, even at $U=\infty$, but there is
a finite local minimum in $U_c$ when the Fermi energy is at the local
maximum in $\epsilon(k)$.

This leads us to the question of what general properties are required in
order to obtain metallic ferromagnetism in this model.
The general picture is that with hole doping, $t_2$ must be less than
zero in order to obtain a ferromagnetic state.
When this condition
is satisfied, the ferromagnetic state occurs over a wide range of
parameters, with, in some cases, quite small $U_c$.
The mechanism for ferromagnetism can be motivated from a local point
of view, in that when $t_2$ is negative the triangular 
structure of the chain frustrates the  antiferromagnetic order (a
generic effect for lattice models).
That this frustration can lead to a ferromagnetic ground state can be
seen on small cluster calculations.

Another point of view emphasizes the importance of the form of the
single-particle density of states.
Wahle et al., \cite{Vollhardt97} for example, emphasize that a
necessary condition for ferromagnetism is an asymmetric density of
states, with a strong singularity and a larger density of states in the
lower part of the band.
Our results here also support these ideas.
For $t_2 = 0$, the density of states is symmetric and there is no
ferromagnetism.
When $t_2 < 0$ the density of states becomes
asymmetric, with the $\epsilon^{-1/2}$ Van Hove singularity at the
lower band edge gaining in weight.
For $t_2 < -0.25$, the presence of the additional Van Hove singularity
at the ferromagnetic Fermi level further stabilizes the ferromagnetic
ground state.

The weak-coupling behavior of this model has also proven to be quite
interesting. 
For weak negative $t_2$, the low-energy effective behavior of the
model does not differ qualitatively from that of the one-dimensional
Hubbard model.
For weak $U$, we do indeed find that the model is well-described as a
Luttinger liquid, and have been able to extract the
Luttinger-liquid parameters, $K_{\rho}, u_{\rho}$ and $u_{\sigma}$
using the DMRG.
However, unlike the one-dimensional Hubbard model, the Luttinger
liquid state of the $t_1$-$t_2$ model undergoes a transition to a
ferromagnetic state at finite $U$.
While it is clear that the breakdown of the Luttinger liquid is not
described within the usual weak-coupling picture, we have tried to
indicate how the breakdown occurs within this picture.
The spin velocity, which goes to zero asymptotically as
$U\rightarrow\infty$ in the $t_2=0$ case, becomes zero at finite $U_c$
for $t_2 < 0$.
The ferromagnetic susceptibility, which is inversely proportional to
the spin velocity in a Luttinger liquid thus diverges at the
transition, implying that the transition is second order.
By fitting this susceptibility with the form $|U-U_c|^{-\gamma}$, we
obtain a critical exponent $\gamma = 2.0 \pm 0.1$.

In addition, we have calculated the spin-spin correlation function in
the vicinity of the transition and find that it becomes ferromagnetic
and exponentially decaying just below the transition.
The correlation length grows as the transition is approached from
below, which is consistent with a second order transition.
We have attempted to extract a critical exponent $\nu$ by fitting the
correlation length to a form $\xi \sim |U-U_c|^{-\nu}$, but find that
it is difficult to extract an exponent due to convergence problems
which limit the maximum lattice size near the transition.

An examination of the behavior of the ground state energy as function
of $U$ near the transition using exact diagonalization suggests
that $E_0(U)$ becomes smooth in the thermodynamic limit and provides
further evidence that the transition is second order.

Finally, we have investigated the very rich low-$U$ phase diagram. 
For a large region of parameters the low-$U$ phase diagram of the
$t_1-t_2$ model can be mapped to that of the  two-chain Hubbard model. 
For sufficiently large $|t_2|$ and a wide range of $n<1$, we confirm
numerically the existence of the doped spin-liquid phase (C1S0)
predicted by weak coupling RG, which is the one-dimensional analog of
a superconducting phase. 
In addition, we have presented evidence for the existence of a new
2 $\times$ C1S1 phase (not found in the weak coupling treatment) in a
region in which we think that the $t_1-t_2$ model behaves as two
uncoupled Hubbard chains.
Because we have also found some additional discrepancies between the
weak-coupling phase diagram and the numerical calculations, more work
needs to be done to clarify the details of the phase diagram for the
lattice model.


\section*{Acknowledgments}

This work was supported by the Swiss National Foundation under the Grant
No. 20-46918.96.
We would like to thank D.\ Baeriswyl, M.\ Dzierazawa,
and J.\ Voit for helpful discussions.


\section{TABLES}

\begin{table}
\[
\begin{array}{|crrrrr|} \hline
C_{4v} \;\;\; & E & 2C_4 & C_2 & 2\sigma_v & 2\sigma_d \\ \hline
A_1  &  1  &  1  &  1  &  1  &  1  \\
A_2  &  1  &  1  &  1  & -1  & -1  \\
B_1  &  1  & -1  &  1  &  1  & -1  \\
B_2  &  1  & -1  &  1  & -1  &  1  \\
E    &  2  &  0  & -2  &  0  &  0  \\ \hline
\end{array} 
\]
\caption{Character table of the group $C_{4v}$.}
\label{tablec4v}
\end{table}

\begin{table}
\[
\begin{array}{|lcc|lc|} \hline
A_1 : & -2t_2 + \sqrt{4t_2^2+8t_1^2} & & E : & -2t_1   \\
     & -2t_2 - \sqrt{4t_2^2+8t_1^2} & &  &  -2t_1 \\
A_2 : &  4t_2 & & &  0 \\
B_1 : &  4t_2 & & &  0 \\
B_2 : & -4t_2 & & &  2t_1 \\
     &     0 & & &  2t_1\\ \hline
\end{array}
\]
\caption{Eigenvalues of the square model for $N=2$.}
\label{tableN2}
\end{table}

\begin{table}
\[
\begin{array}{|lcc|lc|} \hline
A_1 : &   t_1 - 2t_2  & & E : & -2t_2  \\
A_2 : & -2t_1 + 2t_2  & &     &  -2t_2 \\ 
      &   t_1 + 2t_2  & &     &  -\sqrt{3t_1^2+4t_2^2} \\
B_1 : &  2t_1 + 2t_2  & &     &  -\sqrt{3t_1^2+4t_2^2} \\
      &  -t_1 + 2t_2  & &     &  \sqrt{3t_1^2+4t_2^2} \\
B_2 : &  -t_1 - 2t_2  & &     &  \sqrt{3t_1^2+4t_2^2} \\ \hline
\end{array}
\]
\caption{Eigenvalues of the square model for $N=3$.}
\label{tableN3}
\end{table}


\appendix
\onecolumn
\section{EDWARDS ANSATZ}
\label{gradient}

The gradient of $E$ with respect to the one-particle orbitals 
$\varphi_{\alpha}(j)=\varphi_{\alpha j}$ can for real wave functions be 
simplified to
\begin{equation}
  \frac{\partial E}{\partial\varphi_{\alpha j}} = 2 \langle \chi |(H-E)
    \frac{\partial}{\partial\varphi_{\alpha j}} | \chi \rangle .
\end{equation}
This expression can be evaluated using Wick's theorem, yielding
\begin{equation}
  \frac{\partial E}{\partial\varphi_{\alpha j}} = F^{\downarrow}_{\alpha j}
+ F^{\uparrow}_{\alpha j} + F^{U}_{\alpha j}
\end{equation}
where
\begin{eqnarray}
   F^{\downarrow}_{\alpha j} &=& -t_1 \cos (q) \det S^{(1)} \left\{ \
\sum_{\beta}   \varphi_{\beta j+1} [S^{(1)^{-1}}]_{\beta\alpha} +
  \varphi_{\beta j-1} [S^{(1)^{-1}}]_{\alpha\beta} - 2\varphi_{\alpha j}
\right\}    \nonumber \\
&& -t_2 \cos(2q) \det S^{(2)} \left\{ \sum_{\beta}
  \varphi_{\beta j+2} [S^{(2)^{-1}}]_{\beta\alpha} +
  \varphi_{\beta j-2} [S^{(2)^{-1}}]_{\alpha\beta} - 2\varphi_{\alpha j}
 \right\} \nonumber \\
 F^{\uparrow}_{\alpha j} &=& -t_1 \cos(q)  \left\{ \sum_{\beta} 
  \varphi_{\beta j}
  \left( S^{(1)}_{\alpha\beta} + S^{(1)}_{\alpha\beta} \right) 
  - \varphi_{\alpha j+1} - \varphi_{\alpha j-1} \right\} \nonumber \\
  && - t_2 \cos(2q)  \left\{ \sum_{\beta} \varphi_{\beta j}
  \left( S^{(2)}_{\alpha\beta} + S^{(2)}_{\alpha\beta} \right) 
  - \varphi_{\alpha j+2} - \varphi_{\alpha j-2} \right\} \nonumber \\
   F^{U}_{\alpha j} &=& U  \varphi_{\alpha 0} \left\{ \delta_{j0} -
  \sum_{\beta} \varphi_{\beta 0} \varphi_{\beta j}  \right\} \nonumber
\end{eqnarray}
with overlap matrices $ S_{\alpha \beta}^{(i)} $ defined as
\begin{equation}
   S_{\alpha \beta}^{(i)} = \sum_{\ell=1}^{L} \varphi_{\alpha}^{*}(\ell)
         \varphi_{\beta}(\ell+i) .
\end{equation}

\twocolumn



\begin{thebibliography}{99}
\bibitem{spinnote} There are other definitions which actually converge
better with $L$ to the $L=\infty$ result. 
Indeed, for $t_2=0$ Bethe Ansatz calculations \cite{Beduerftig} show
that $E_1(S_z=1)-E_0(S_z=1)$, where $E_1$ is the first  
spin-particle-hole-excitation, converges much faster with system size
than Eq.\ (\ref{eqspingapdef}).
\end{thebibliography}
\end{document}